# Concentric UCAs Based Low-Order OAM for High Capacity in Radio Vortex Wireless Communications

Haiyue Jing, *Student Member, IEEE*, Wenchi Cheng, *Senior Member, IEEE*,
Zan Li, *Senior Member, IEEE*, and Hailin Zhang, *Member, IEEE*

**Abstract**

Due to the potential capacity-boosting for wireless communications, the Radio vOrtex Wireless COMMunication (RowComm) over orthogonal states/modes of Orbital Angular Momentum (OAM) has been paid much attention in recent years. Uniform circular array (UCA), as an efficient and convenient antenna structure, can transmit/receive multiple OAM beams with different OAM-modes simultaneously when the transmitter and receiver are aligned. However, for high-order OAM-modes, the OAM beams are divergent accompanied by severe attenuations. Thus, it is difficult to directly use high-order OAM-modes to achieve high capacity for RowComms. To obtain high capacity potentially offered by OAM-modes, in this paper we transform the singular UCA into the concentric UCAs, where high capacity can be achieved using multiple parallel low-order OAM-modes instead of all high-order OAM-modes, to increase the capacity of transmitter-receiver aligned RowComms. In particular, we study two cases: 1). Concentric UCAs based RowComms without co-mode interference; 2). Concentric UCAs based RowComms with co-mode interference. We propose a co-mode-interference-free and a co-mode-interference-contained mode-decomposition schemes to recover transmit signals for these two cases, respectively. Also, we develop the optimal power allocation schemes to maximize the capacity for these two cases. Numerical simulations are presented to validate and evaluate that our developed concentric UCAs based low-order RowComms can significantly increase the capacity as compared with that of singular UCA based RowComms.

The authors are with the State Key Laboratory of Integrated Services Networks, Xidian University, Xi'an, 710071, China (e-mails: hyjing@stu.xidian.edu.cn, wccheng@xidian.edu.cn, zanli@xidian.edu.cn, and hlzhang@xidian.edu.cn).

# Concentric UCAs Based Low-Order OAM for High Capacity in Radio Vortex Wireless Communications

## I. Introduction

As wireless communications move from the fourth-generation (4G) to the fifth-generation (5G) and the future 5G-beyond, it is urgent to meet the significantly increasing capacity demand [1], [2]. However, not only the widely used traditional wireless transmission techniques such as orthogonal frequency-division multiplexing (OFDM) [3], but also the recent developed new techniques such as co-frequency co-time full-duplex [4] and non-orthogonal multiplexing [5] cannot achieve the boosting capacity requirements for wireless communications. In other words, it is now very difficult to use the traditional plane-electromagnetic (PE) wave based wireless communications to satisfy the ever-growing demand for capacity.

Fortunately, the electromagnetic (EM) wave possesses not only linear momentum, but also angular momentum, which includes the spin angular momentum (SAM) and Orbital Angular Momentum (OAM) [6]–[11]. The orbital angular momentum (OAM), which is a kind of wave-front with helical phase and has not been well studied yet, is another important property of EM wave [6]–[8]. The OAM-based vorticose wave has different topological charges, which are orthogonal with each other, bridging a new way to significantly increase the capacity of wireless communications [9].

There exist some experiments that have verified the feasibility of OAM based Radio vOrtex Wireless COMMunications (RowComms) [6], [12]–[18]. The authors of [13], [14] studied the OAM based RowComms with two OAM-modes (OAM-modes $0$ and $1$), which share the same frequency band. The authors of [15], [16] performed the OAM based high capacity transmission with $60$ GHz and $17$ GHz carrier frequencies, respectively. The authors of [19] experimentally evaluated patch antenna arrays for OAM generation at 60 GHz. The authors of [20], [21] demonstrated OAM multiplexing can achieve high capacity in millimetre-wave communications. Also, OAM based RowComms have received much attention in the aspect of mode detection [22], [23], mode separation [24], axis estimation and alignment [25], mode modulation [26], OAM-beams converging [27], and mode hopping [28], etc.

For high capacity wireless communications, multiple OAM-modes need to be generated and used simultaneously. Uniform circular array (UCA), which is flexible in radiating multiple OAM beams with different charge numbers [29], is one of intriguing antenna structures for multiple OAM-modes based RowComms [30], [31]. The authors of [32] showed that UCA is better than the radial array and the tangential array to generate the desired OAM beam with linear excitation. Noticing that the array-elements of UCA are fed with the same input signal [6], the OAM signal can be generated within one RF-chain based antenna which has several array-elements [33]. It is concluded that for a UCA with a limited number of array-elements represented by $U$, there exists the limitation on the largest OAM number $l$ $\left[(1-U)/2 \leq l \leq U/2\right]$, where $l$ represents the number of topological charges, i.e., the number of OAM-modes [10]. However, not all OAM-modes can be directly used to achieve the maximum capacity for RowComms. The beams for high-order OAM-modes (the absolute value of $l$ is relatively large) are divergent and experience severe attenuation in free space [27], [29]. Thus, the received signal-to-noise ratio (SNR) is relatively low for high-order OAM-modes, which severely decreases the capacity of OAM based RowComms. The authors of [27] designed the lens based converging for the OAM-modes, which can increase the capacity. However, the divergent angle for converged high-order OAM-modes using the lens converging is still relatively high, i.e., capacities related to high-order OAM-modes are smaller than those related to low-order OAM-modes.

To solve the above-mentioned problem, in this paper we transform the singular UCA based RowComm model into an equivalent concentric UCAs based RowComm model, where different UCAs have the same center but different radii and the transmitter and receiver are aligned with each other. In our proposed concentric UCAs based RowComms model, we assume zero mutual coupling among array-elements within one UCA antenna. In terms of mutual coupling between two adjacent concentric UCAs, we consider the mutual coupling impacted correlation among concentric UCAs, i.e., if there exists mutual coupling between two adjacent concentric UCAs, the two concentric UCAs are correlated. While if there is no mutual coupling between two adjacent concentric UCAs, the two concentric UCAs are uncorrelated. For the concentric UCAs based RowComm model, “co-mode” represents the OAM-modes which have the same order corresponding to different concentric UCAs. If all concentric UCA antennas (One UCA antenna consists of the array-elements with the same radius) are uncorrelated (“uncorrelated” represents that there is no co-mode interference among all UCA antennas), there is no co-mode interference among all UCA antennas. Otherwise, if all concentric UCA antennas are correlated (“correlated” represents that there exists co-mode interference among all UCA antennas), there exists co-mode interference among all UCA antennas. Thus, we consider the co-mode-interference-free and the co-mode-interference-contained models, respectively, which are discussed in the following.

- For the co-mode-interference-free model, we develop a mode-decomposition scheme and a multiplexing-detection scheme to obtain the signal corresponding to each OAM-mode of each transmit UCA. Then, we develop the optimal power allocation scheme to obtain the maximum capacity for the concentric UCAs based RowComms without co-mode interference.
- For the co-mode-interference-contained model, we propose a mode-decomposition scheme to decompose the OAM-modes and a co-mode zero-forcing successive interference cancellation (CM-ZF-SIC) algorithm, which is used to recover the signals carried by co-modes corresponding to the transmit concentric UCAs. Then, we formulate the capacity maximization problem and convert the capacity maximization problem to an equivalent convex optimization problem. We derive the optimal power allocation scheme to maximize the capacity of RowComms with co-mode interference.

Also presented are the numerical simulation results which validate our developed schemes, showing that concentric UCAs based RowComms can significantly increase the capacity as compared with the singular UCA based RowComms.

The rest of this paper is organized as follows. Section II gives the concentric UCAs based RowComms model. Section III proposes two mode-decomposition schemes to obtain the signal on each OAM-mode of each transmit UCA-antenna. Section IV develops the optimal power allocation schemes to maximize the capacity of the concentric UCAs based RowComms and answers the question why the capacity of the concentric UCAs based RowComms is larger than the capacity of the singular UCA based RowComms. Section V evaluates our developed schemes and compares the capacity of the concentric UCAs based RowComms with the capacity of the singular UCA based RowComms. Section VI concludes this paper.

## II. UCA Based System Model for RowComms

Figure 1 depicts the system models for the singular UCA (Fig. 1(a)) and concentric UCAs (Fig. 1(b)) based RowComms, where the transmitter and the receiver are aligned with each other and the sizes of them can be different. $d$ denotes the distance from the center of the transmit UCA to the center of the receive UCA. For the transmit UCA, the array-elements, which are

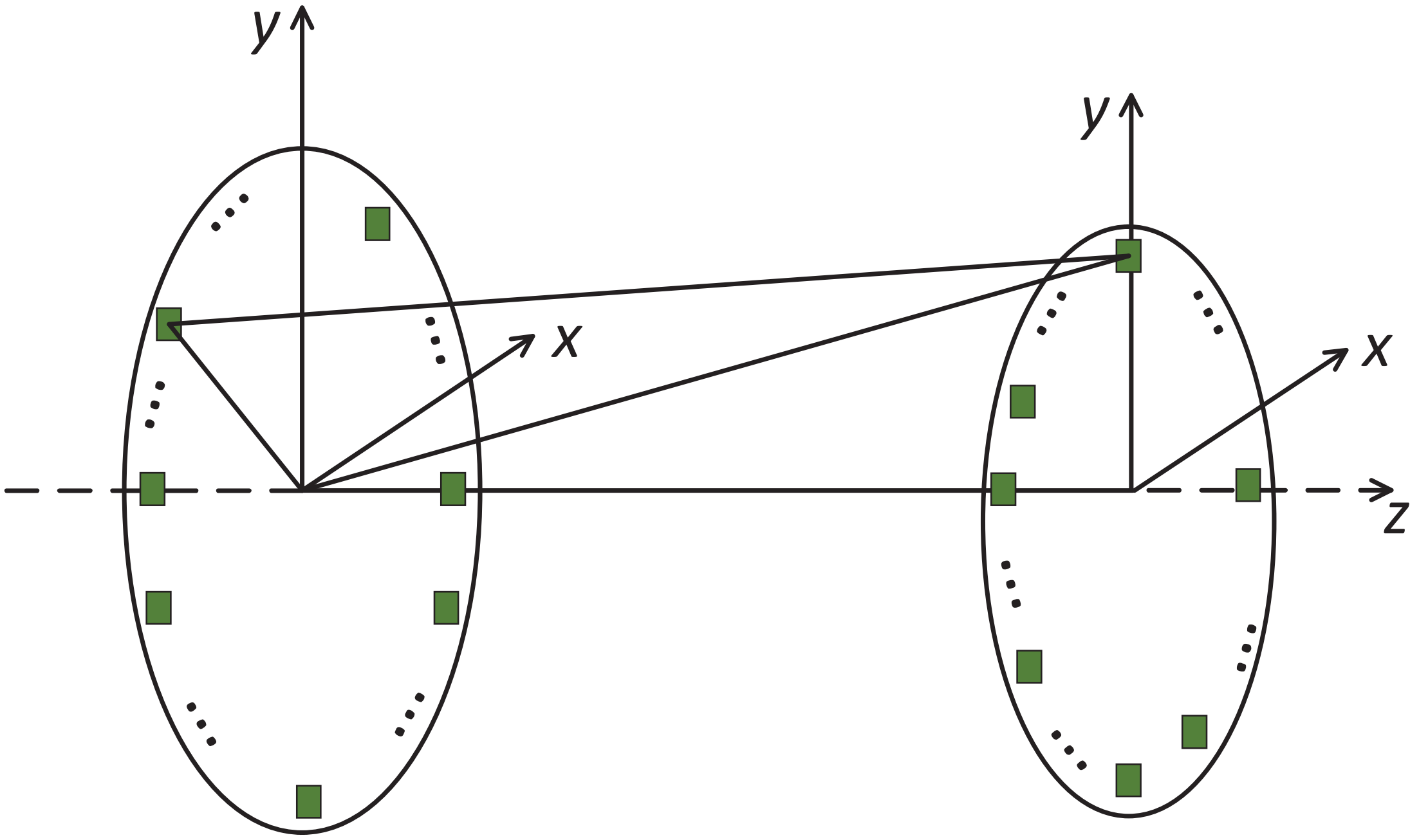

(a) The system model for the singular UCA based RowComm .

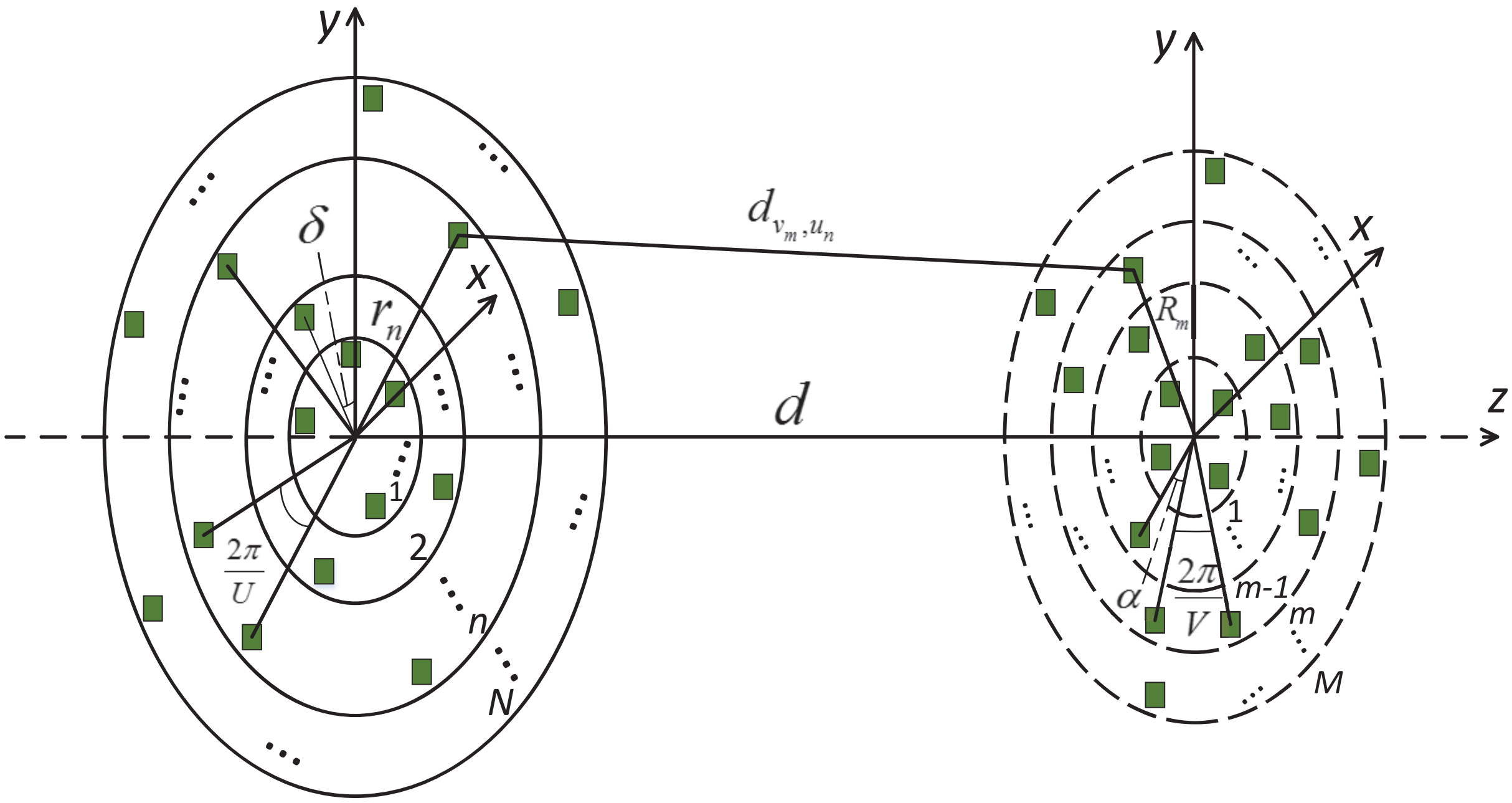

(b) The system model for the concentric UCAs based RowComm.

Fig. 1. UCA based system model for RowComms.

fed with the same input signal but with a successive delay from array-element to array-element such that after a full turn the phase has been incremented by an integer multiple $l$ of $2\pi$, are uniformly around the perimeter of the circle. For the receive UCA, the array-elements are also uniformly around perimeter of the circle.

Figure 1(a) shows the singular UCA based RowComms system model, where the transmit and receive antennas are both a singular UCA. It is expected that the capacity increases as the number of OAM-modes increases. However, high-order OAM beams experience severe attenuation resulting in very low received SNR at the receive UCA. Therefore, it is very difficult to obtain high capacity by directly using the high-order OAM-modes in RowComms.

There are plenty of area unused within the circles of transmit UCA and receive UCA. Thus, we can fully make use of the unused area to equip more UCAs to exploit new low-order OAM-modes for high capacity communications, as shown in Fig. 1(b). In Fig. 1(b), there are $N$ concentric transmit UCAs and $M$ concentric receive UCAs, where each transmit UCA is equipped with $U$ ($1 \leq u \leq U$) array-elements on a circle and each receive UCA is equipped with $V$ ($1 \leq v \leq V$) array-elements on a circle. We assume zero mutual coupling among array-elements within one UCA antenna in this paper. If mutual coupling among array-elements within one UCA antenna is incorporated, the wavefront of OAM-mode changes, resulting that the orthogonality among OAM-modes doesn't strictly hold. Thus, the wavefront phase change corresponding to each OAM-mode needs to be estimated and it makes trouble for the receiver to decompose OAM beams with multiple OAM-modes. When the distance between two adjacent concentric UCAs is larger than 0.5 $\lambda$, we consider there is zero mutual coupling among concentric UCAs and all transmit UCAs are uncorrelated, i.e., there is no co-mode interference among the same order of OAM-modes corresponding to different UCAs. Thus, co-mode-interference-free model can be used for this case. When the distance between two adjacent concentric UCAs is smaller than 0.5 $\lambda$, there exists mutual coupling among concentric UCAs and all transmit UCAs are correlated, i.e., there exists co-mode interference among the same order of OAM-modes corresponding to different UCAs. Thus, co-mode-interference-contained model can be used for this scenario. We denote by $\delta$ and $\alpha$ the angles between two adjacent array-elements for transmit UCAs and receives UCAs, respectively. Also, we denote by $r_n$ the radius of the $n$th transmit UCA and $R_m$ the radius of the $m$th receive UCA. In addition, we denote by $d_{v_m,u_n}$ the distance between the $v$th array-element on the $m$th receive UCA and the $u$th array-element on the $n$th transmit UCA. In the following, we develop the mode-decomposition scheme, the multiplexing-detection scheme, and the optimal power allocation scheme for the co-mode-interference-free model based RowComms. We also develop the mode-decomposition scheme, the CM-ZF-SIC algorithm, and the optimal power allocation scheme for the co-mode-interference-contained model based RowComms.

## III. Mode-Decomposition for Concentric UCAs Based RowComms

The signal at the $u$th array-element on the $n$th transmit UCA, denoted by $x_{n,u}$, is given as follows:

$$x_{n,u} = \sum_{l=\frac{1-U}{2}}^{U/2} \frac{1}{\sqrt{U}} s_{n,l} e^{j\varphi_{u_n} l} = \sum_{l=\frac{1-U}{2}}^{U/2} \frac{1}{\sqrt{U}} s_{n,l} e^{j(\phi_u + n\delta)l} = \sum_{l=\frac{1-U}{2}}^{U/2} \frac{1}{\sqrt{U}} s_{n,l} e^{j\left[\frac{2\pi(u-1)}{U} + \frac{2\pi n}{NU}\right]l}, \tag{1}$$

where $l$ is the OAM-mode number (also the order of OAM-mode) and $s_{n,l}$ is the signal on the $l$th OAM-mode of the $n$th transmit UCA. The azimuthal angle, defined as the angular position on a plane perpendicular to the axis of propagation, is given by $\varphi_{u_n} = \phi_u + n\delta$ corresponding to the $u$th array-element on the $n$th transmit UCA, where $u_n$ is the $u$th array-element of the $n$th transmit UCA, $\phi_u = 2\pi(u-1)/U$ is the basic angle for each transmit UCA, and $\delta = 2\pi/NU$ is the minimal interval for the angle of rotation corresponding to the transmit UCA. The transmit signal, denoted by $\widetilde{x}_{n,l}$, of the $l$th OAM-mode from the $n$th transmit UCA can be treated as a continuous signal expressed as follows:

$$\widetilde{x}_{n,l} = s_{n,l} e^{j\varphi l}, \tag{2}$$

where $\varphi$ $(0 \leq \varphi < 2\pi)$ is the continuous azimuthal angle. Here, the discrete azimuthal angle in terms of $u$ turns to be the continuous azimuthal angle in terms of $\varphi$ along with the OAM signal transmitting (This is the spatial Digital to Analog (DA) process, which is defined as the process that the discrete OAM signal is converted to be the continuous OAM signal through the UCA antenna.).

We denote by $h_{v_m,u_n}$ the channel gain from the $u$th array-element on the $n$th transmit UCA to the $v$th array-element on the $m$th receive UCA. Then, $h_{v_m,u_n}$ can be written as follows [10]:

$$h_{v_m,u_n} = \frac{\beta\lambda e^{-j\frac{2\pi}{\lambda} d_{v_m,u_n}}}{4\pi d_{v_m,u_n}}, \tag{3}$$

where $v_m$ represents the $v$th array-element on the $m$th receive UCA. The parameter $\beta$ denotes all relevant constants such as attenuation and phase rotation caused by antennas and their patterns on both sides. We denote by $s$ the distance between the projection of the $u$th array-element on the $n$th transmit UCA in the receive plane and the $v$th array-element on the $m$th receive UCA. Then, we have

$$s = \sqrt{r_n^2 + R_m^2 - 2r_n R_m \cos\left(\phi_u + n\delta - \psi_v - m\alpha\right)}, \tag{4}$$

where $\psi_v = 2\pi(v-1)/V$ is the basic angle for each receive UCA, $\alpha = 2\pi/MV$ is the minimal interval for the angle of rotation corresponding to the receive UCAs, and $\psi_v + m\alpha$ is the azimuthal angle of the $v$th array-element on the $m$th receive UCA. Then, we can write $d_{v_m,u_n}$ as follows:

$$d_{v_m,u_n} = \sqrt{d^2 + s^2} = \sqrt{d^2 + r_n^2 + R_m^2 - 2r_n R_m \cos\left(\phi_u + n\delta - \psi_v - m\alpha\right)}. \tag{5}$$

Because of $d_{v_m,u_n} \gg r_n$ and $d_{v_m,u_n} \gg R_m$, we can make approximation for $d_{v_m,u_n}$ at the denominator and numerator of Eq. (3). For the denominator, we use $d_{v_m,u_n} \approx d$. For the numerator, $d_{v_m,u_n}$ can be rewritten according to $\sqrt{1-x} \approx 1 - x/2$ as follows:

$$\begin{aligned} d_{v_m,u_n} &= \sqrt{d^2 + r_n^2 + R_m^2}\sqrt{1 - \frac{2r_n R_m \cos\left(\phi_u - \psi_v + n\delta - m\alpha\right)}{d^2 + r_n^2 + R_m^2}} \\ &\approx \sqrt{d^2 + r_n^2 + R_m^2} - \frac{r_n R_m \cos\left(\phi_u - \psi_v + n\delta - m\alpha\right)}{\sqrt{d^2 + r_n^2 + R_m^2}}. \end{aligned} \tag{6}$$

Then, using the above-mentioned approximation, we can rewrite $h_{v_m,u_n}$ as follows:

$$h_{v_m,u_n} = \frac{\beta\lambda e^{-j\frac{2\pi}{\lambda}\sqrt{d^2+r_n^2+R_m^2}}}{4\pi d} \exp\left[\frac{j2\pi r_n R_m}{\lambda\sqrt{d^2 + r_n^2 + R_m^2}} \cos\left(\phi_u - \psi_v + n\delta - m\alpha\right)\right]. \tag{7}$$

Before spatially sampling at the receiver, the signal received at the $v$th array-element of the $m$th receive UCA, denoted by $r_{m,v}$, can be derived as follows:

$$r_{m,v} = \sum_{l=\frac{1-U}{2}}^{U/2} \sum_{n=1}^{N} \sum_{u=1}^{U} h_{v_m,u_n} \frac{1}{\sqrt{U}} s_{n,l} e^{j\left[\frac{2\pi(u-1)}{U} + n\delta\right]l} = \sum_{l=\frac{1-U}{2}}^{U/2} \sum_{n=1}^{N} h_{v_m,n,l} s_{n,l}, \tag{8}$$

where $h_{v_m,n,l}$ is the channel gain for the $l$th OAM-mode corresponding to the $n$th transmit UCA and the $v$th array-element on the $m$th receive UCA, and can be given as follows:

$$h_{v_m,n,l} = \sum_{u=1}^{U} \frac{\beta\lambda e^{-j\frac{2\pi}{\lambda}\sqrt{d^2+r_n^2+R_m^2}}}{4\pi d\sqrt{U}} e^{j\left[\frac{2\pi(u-1)}{U} + n\delta\right]l} \exp\left[\frac{j2\pi r_n R_m}{\lambda\sqrt{d^2 + r_n^2 + R_m^2}} \cos\left(\phi_u - \psi_v + n\delta - m\alpha\right)\right]. \tag{9}$$

For convenient expression, we denote by $\vartheta = \psi_v - n\delta + m\alpha$. Then, we have

$$\begin{aligned}
&\frac{j^l}{U}\sum_{u=1}^{U}\exp\left[j\left(\phi_u-\vartheta\right)l\right]\exp\left[\frac{j2\pi r_n R_m}{\lambda\sqrt{d^2+r_n^2+R_m^2}}\cos\left(\phi_u-\vartheta\right)\right]\\
&\quad\approx \frac{j^l}{2\pi}\int_0^{2\pi}\exp\left[j\left(\widetilde{\phi}_u-\vartheta\right)l\right]\exp\left[\frac{j2\pi r_n R_m}{\lambda\sqrt{d^2+r_n^2+R_m^2}}\cos\left(\widetilde{\phi}_u-\vartheta\right)\right]d\widetilde{\phi}_u\\
&\quad= \frac{j^l}{2\pi}\int_{-\vartheta}^{2\pi-\vartheta}\exp\left[j\left(\widetilde{\phi}_u-\vartheta\right)l\right]\exp\left[\frac{j2\pi r_n R_m}{\lambda\sqrt{d^2+r_n^2+R_m^2}}\cos\left(\widetilde{\phi}_u-\vartheta\right)\right]d(\widetilde{\phi}_u-\vartheta)\\
&\quad= J_l\left(\frac{2\pi r_n R_m}{\lambda\sqrt{d^2+r_n^2+R_m^2}}\right),
\end{aligned}\tag{10}$$

where $\phi_u = 2\pi(u-1)/U$ and $\widetilde{\phi}_u$ is a continuous variable of $\phi_u$ ranging from 0 to $2\pi$, since when $u$ is equal to 1, $\phi_u$ is zero and when $u = U$, $\phi_u$ is $2\pi - 2\pi/U$ and very close to $2\pi$. The approximation error, denoted by $Er$, between the Bessel function and the sum of polynomials in Eq. (10) is given as follows:

$$Er{=}J_l\left(\frac{2\pi r_n R_m}{\lambda\sqrt{d^2+r_n^2+R_m^2}}\right)-\frac{j^l}{U}\sum_{u=1}^{U}\exp\left[j\left(\phi_u-\vartheta\right)l\right]\exp\left[\frac{j2\pi r_n R_m}{\lambda\sqrt{d^2+r_n^2+R_m^2}}\cos\left(\phi_u-\vartheta\right)\right].\tag{11}$$

Figure 2 shows the approximation error for different OAM-modes. As shown in Fig. 2, we can

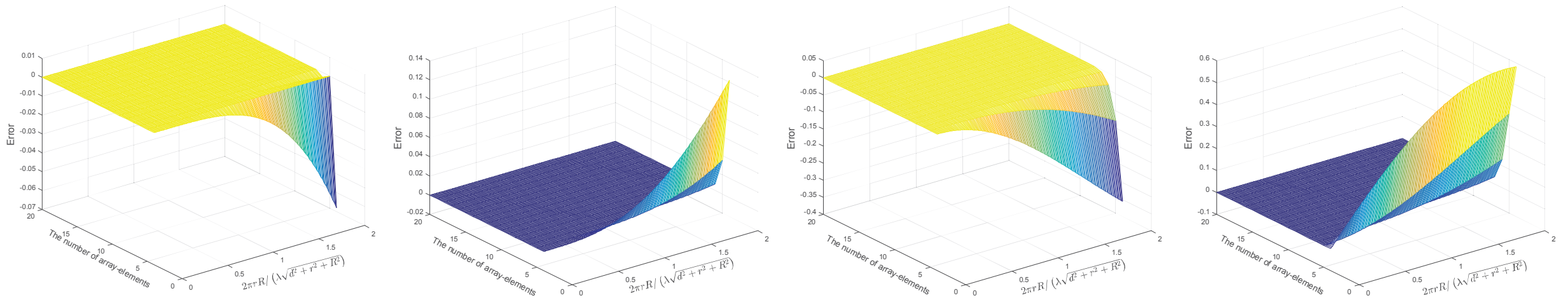

$l = 0$ (PE beam) $\quad l = 1 \quad l = 2 \quad l = 3$

Fig. 2. Approximation errors for different OAM-modes.

obtain that the error is very small when the number of array-elements is larger than 10. Thus, we can neglect the error.

Taking Eq. (10) into Eq. (9), we can approximate $h_{v_m,n,l}$ as follows:

$$h_{v_m,n,l} \approx \frac{\beta\lambda\sqrt{U}e^{-j\frac{2\pi}{\lambda}\sqrt{d^2+r_n^2+R_m^2}}e^{j\left[\frac{2\pi(v-1)}{V}+m\alpha\right]l}}{4\pi d j^l}J_l\left(\frac{2\pi r_n R_m}{\lambda\sqrt{d^2+r_n^2+R_m^2}}\right),\tag{12}$$

where

$$J_l(\alpha) = \frac{j^l}{2\pi} \int_0^{2\pi} e^{jl\tau} e^{j\alpha \cos\tau} d\tau \tag{13}$$

is the $l$-order Bessel function [34]. Eq. (12) shows that for different array-elements on the $m$th receive UCA, the corresponding channel gains $h_{v_m,n,l}$ follow the Bessel distribution with different orders. The azimuthal angle is continuous when the radio vortex signal is transmitted by the transmit UCA. At the receiver, the item $\exp\{j\varphi l\}$ turns to $\exp\{j[2\pi(v-1)/V + m\alpha]l\}$ after spatially sampling. Then, we can obtain all the channel gains from the $n$th transmit UCA to the $m$th receive UCA. We derive the channel gain, denote by $\widetilde{h}_{mn,l}$, for the channel from the $n$th transmit UCA to the $m$th receive UCA as follows:

$$\widetilde{h}_{mn,l} = \frac{\beta\lambda\sqrt{U} e^{-j\frac{2\pi}{\lambda}\sqrt{d^2+r_n^2+R_m^2}} e^{j\varphi l}}{4\pi d j^l} J_l\left(\frac{2\pi r_n R_m}{\lambda\sqrt{d^2+r_n^2+R_m^2}}\right). \tag{14}$$

Observing Eq. (14), we can find that the discrete azimuthal angle in terms of $u$ turns to be the continuous azimuthal angle in terms of $\varphi$ along with the OAM signal transmitting (spatial DA process) and the transmit UCA makes the OAM signal like going through the Bessel-form channel. The channel gains mainly depend on the order of OAM-mode and the value of $2\pi r_n R_m / \left(\lambda\sqrt{d^2+r_n^2+R_m^2}\right)$. The order of OAM-mode represents the order of Bessel function and $2\pi r_n R_m / \left(\lambda\sqrt{d^2+r_n^2+R_m^2}\right)$ represents the independent variable of Bessel function. According to the characteristics of Bessel function, $\widetilde{h}_{mn,l}$ severely decreases as $l$ increases because the value of $2\pi r_n R_m / \left(\lambda\sqrt{d^2+r_n^2+R_m^2}\right)$ is small, resulting in very small received SNR when using the high-order OAM-mode (The OAM-mode number is relatively large). That is not what we expected since we aim to obtain the maximum capacity offered by all orthogonal OAM-modes. In fact, it has been shown that the electromagnetic wave with OAM is vorticose hollow and divergent [20]. Moreover, as the order of OAM-mode increases, the corresponding EM wave becomes more divergent. Therefore, if the order of OAM is relatively larger, it is impossible to get high capacity for the high-order OAM-modes based RowComms since the received SNR on the high-order OAM-mode is very small as compared with the PE wave based wireless communications.

We denote by $h_{mn,l}$ the channel gain without the item $e^{j\varphi l}$ as follows:

$$h_{mn,l} = \frac{\beta\lambda\sqrt{U}e^{-j\frac{2\pi}{\lambda}\sqrt{d^2+r_n^2+R_m^2}}}{4\pi d j^l} J_l\left(\frac{2\pi r_n R_m}{\lambda\sqrt{d^2+r_n^2+R_m^2}}\right). \tag{15}$$

Clearly, $h_{mn,l}$ is independent of $u$ and $v$.

### A. Mode-Decomposition Scheme for Concentric UCAs Based RowComms

The continuous azimuthal angle in terms of $\varphi$ is spatially sampled to be the discrete azimuthal angle in terms of $v$ along with the OAM signal reception (This is the spatial Analog to Digital (AD) process, which is defined as the process that the continuous OAM signal is changed to the discrete OAM signal through the UCA antenna.). The signal received at the $v$th array-element of the $m$th receive UCA, denoted by $y_{m,v}$, can be obtained as follows:

$$y_{m,v} = \sum_{l=\frac{1-U}{2}}^{U/2} \sum_{n=1}^{N} h_{mn,l} s_{n,l} e^{j\left[\frac{2\pi(v-1)}{V}+m\alpha\right]l} + w_{m,v}, \tag{16}$$

where $w_{m,v}$ denotes the received noise at the $v$th array-element of the $m$th receive UCA. The noise is uncorrelated among OAM-modes and among concentric UCAs. To obtain the received signal on the $l_0$th $\left(\frac{1-U}{2} \leq l_0 \leq \frac{U}{2}\right)$ OAM-mode sent from all transmit UCAs, we multiply $y_{m,v}$ with $\exp\{-j[2\pi(v-1)+m\alpha]l_0/V\}$ and we have the received signal, denoted by $y_{m,v,l_0}$, on the $l_0$th OAM-mode corresponding to the $v$th array-element of the $m$th received UCA as follows:

$$\begin{aligned} y_{m,v,l_0} &= y_{m,v} e^{-j\left[\frac{2\pi(v-1)}{V}+m\alpha\right]l_0} \\ &= \sum_{n=1}^{N} \sum_{l=\frac{1-U}{2}, l\neq l_0}^{U/2} h_{mn,l} s_{n,l} e^{j\left[\frac{2\pi(v-1)}{V}+m\alpha\right](l-l_0)} + \sum_{n=1}^{N} h_{mn,l_0} s_{n,l_0} + w_{m,v} e^{-j\left[\frac{2\pi(v-1)}{V}+m\alpha\right]l_0}. \end{aligned} \tag{17}$$

Since $h_{mn,l}$ is independent of $v$ as it can be seen from Eq. (15), the received signal, denoted by $\widetilde{y}_{m,l_0}$, on the $l_0$th OAM-mode of the $m$th receive UCA can be derived as follows:

$$\begin{aligned}\widetilde{y}_{m,l_0} &= \sum_{v=1}^{V} y_{m,v,l_0} \\ &= \sum_{n=1}^{N} \sum_{l=\frac{1-U}{2}, l\neq l_0}^{U/2} h_{mn,l} s_{n,l} \sum_{v=1}^{V} e^{j\left[\frac{2\pi(v-1)}{V}+m\alpha\right](l-l_0)} + \sum_{n=1}^{N} h_{mn,l_0} \sum_{v=1}^{V} s_{n,l_0} + \sum_{v=1}^{V} w_{m,v} e^{-j\left[\frac{2\pi(v-1)}{V}+m\alpha\right]l_0} \\ &= \sum_{n=1}^{N} V h_{mn,l_0} s_{n,l_0} + \widetilde{w}_{m,l_0}, \end{aligned} \tag{18}$$

where $\widetilde{w}_{m,l_0}$ denotes the received noise on the $l_0$th OAM-mode corresponding to the $m$th receive UCA. Now, we have obtained the estimated decomposed signal, specified in Eq. (18), for all OAM-modes.

We denote by $\boldsymbol{y}_l = [\widetilde{y}_{1,l}, \widetilde{y}_{2,l}, \cdots, \widetilde{y}_{M,l}]^{\mathrm{T}}$ and $\boldsymbol{w}_l = [\widetilde{w}_{1,l}, \widetilde{w}_{2,l}, \cdots, \widetilde{w}_{M,l}]^{\mathrm{T}}$ the received signal and noise vectors, respectively, corresponding to the $l$th OAM-mode, where $[\cdot]^{\mathrm{T}}$ represents the transpose operation.

### *B. Multiplexing-Detection Scheme for Concentric UCAs Without Co-Mode Interference Based RowComms*

Using the zero-forcing detection scheme [35], the estimated transmit signal vector, denoted by $\widehat{\boldsymbol{s}}_{\boldsymbol{l}} = [\widehat{s}_{1,l}, \widehat{s}_{2,l}, \cdots, \widehat{s}_{N,l}]^{\mathrm{T}}$, can be derived as follows:

$$\widehat{\boldsymbol{s}}_{\boldsymbol{l}} = (\boldsymbol{H}_l^{\mathrm{H}}\boldsymbol{H}_l)^{-1}\boldsymbol{H}_l^{\mathrm{H}}\boldsymbol{y}_l = \boldsymbol{s}_l + (\boldsymbol{H}_l^{\mathrm{H}}\boldsymbol{H}_l)^{-1}\boldsymbol{H}_l^{\mathrm{H}}\boldsymbol{w}_l, \tag{19}$$

where $(\cdot)^{\mathrm{H}}$ denotes the conjugation operation, $\boldsymbol{w}_l$ is the received noise vector corresponding to the $l$th OAM-mode, and

$$\boldsymbol{H}_l = V \times \begin{bmatrix} h_{11,l} & h_{12,l} & \cdots & h_{1N,l} \\ h_{21,l} & h_{22,l} & \cdots & h_{2N,l} \\ \vdots & \vdots & \ddots & \vdots \\ h_{M1,l} & h_{M2,l} & \cdots & h_{MN,l} \end{bmatrix}, \tag{20}$$

is the channel gain matrix corresponding to the $l$th OAM-mode. Then, the SNR, denoted by $\mathrm{SNR}_l$, for the $l$th OAM-mode can be derived as follows:

$$\mathrm{SNR}_l = \frac{|\boldsymbol{s}_l|^2}{\sigma_l^2 |(\boldsymbol{H}_l^{\mathrm{H}} \boldsymbol{H}_l)^{-1} \boldsymbol{H}_l^{\mathrm{H}}|^2}, \tag{21}$$

where $\sigma_l^2$ denotes the variance of received noise corresponding to $\widetilde{w}_{m,l}$ for the $l$th OAM-mode of each transmit UCA. Note that the variances of received noise corresponding to different circles are the same when the order of OAM-modes are the same.

### *C. CM-ZF-SIC Algorithm for Concentric UCAs With Co-Mode Interference Based RowComms*

For concentric UCAs based RowComms system, we develop the CM-ZF-SIC algorithm to recover transmit signals based on Eq. (18). The transmit signal is detected in order from large to small in accordance with signal-to-interference-plus-noise ratio (SINR) until all signals are recovered [36], [37].

---

**Algorithm 1** : The CM-ZF-SIC Algorithm

---

**For** $l = \frac{1-U}{2}$ to $\frac{U}{2}$ **do**
1) **Initialization**
 a) $i$=1
 b) $\mathrm{w}_1^l = \boldsymbol{H}_l^{\dagger} = (\boldsymbol{H}_l^{\mathrm{H}} \boldsymbol{H}_l)^{-1} \boldsymbol{H}_l^{\mathrm{H}}$, $\boldsymbol{y} = \boldsymbol{y}_l$
2) **While** $(i \leq N)$
 a) $k_i = \arg\min_j \|(\mathrm{w}_i^l)_\varrho\|^2$
 b) $r_{k_i} = (\mathrm{w}_i^l)_{k_i}^{\mathrm{H}} \boldsymbol{y}_i$
 c) $(\widehat{\boldsymbol{s}}_l)_{k_i} = Q(r_{k_i})$
 d) $\boldsymbol{y}_{i+1} = \boldsymbol{y}_i - (\widehat{\boldsymbol{s}}_l)_{k_i} (\boldsymbol{H}_l)_{k_i}$
 e) $\mathrm{w}_{i+1}^l = \left((\boldsymbol{H}_l)_{\overline{k}_i}\right)^{\dagger}$
 f) $i = i + 1$
  **End while**
**End for**

---

The CM-ZF-SIC algorithm is given by **Algorithm 1**, where the notation $(\cdot)^{\dagger}$ denotes the Moore-Penrose pseudoinverse. For convenient expression, we denote by $\boldsymbol{y}$ a vector which has the same size as $\boldsymbol{y}_l$. The parameter $\mathrm{w}_i^l$ denotes the $i$th matrix corresponding to the $l$th OAM-mode and $\boldsymbol{y}_i$ represents the $i$th matrix. The notation $(\mathrm{w}_i^l)_\varrho$ denotes the $\varrho$th row of the matrix $\mathrm{w}_i^l$ and $k_i$ represents the $\varrho$th row of the matrix $\mathrm{w}_i^l$, which has the minimum $l_2$-norm. The $k_i$th estimated signal is denoted by $r_{k_i}$ and $Q(\cdot)$ is the quantization (slicing) operation. The notation $(\boldsymbol{H}_l)_{k_i}$ represents the $k_i$th column of the matrix $\boldsymbol{H}_l$ while the notation $(\boldsymbol{H}_l)_{\overline{k}_i}$ denotes the matrix obtained by zeroing the elements of columns from $k_1$ to $k_i$ of the matrix.

According to the $l_2$-norm of $\mathbf{w}_i^l$, we can obtain the transmit signals in order for a certain OAM-mode $l_0$. After the cancellation, the residual co-mode interference corresponding to cancelled concentric UCAs is very small as compared with the power of co-mode interfering signals corresponding to other concentric UCAs. Thus, we can ignore the residual cancelled co-mode interference. Then, for the $l$th OAM-mode, the SINR of the $k_i$th signal, denoted by $\mathrm{SINR}_{i,l}$, can be written as follows:

$$\mathrm{SINR}_{i,l} = \begin{cases} \dfrac{\frac{P_{i,l}}{\|(\mathbf{w}_i^l)_{k_i}\|^2}}{\sigma_l^2 + \sum\limits_{\varrho \notin \mathcal{K}_i^l} \frac{P_{\varrho,l}}{\|(\mathbf{w}_i^l)_{\varrho}\|^2}}, & 1 \leq i < N; \\ \dfrac{P_{N,l}}{\sigma_l^2 \|(\mathbf{w}_N^l)_{k_N}\|^2}, & i = N, \end{cases} \tag{22}$$

where $\mathcal{K}_i^l = \{k_1, k_2, \cdots, k_i\}$ denotes the index set of estimated transmit signals corresponding to the transmit signals $(s_{k_1,l}, s_{k_2,l}, \cdots, s_{k_i,l})$ of the $l$th OAM-mode.

## IV. Optimal Power Allocations for Co-Mode-Interference-Free and Co-Mode-Interference-Contained Models

In this section, we develop the optimal power allocation schemes to achieve the maximum capacities for co-mode-interference-free and co-mode-interference-contained models based Row-Comms, respectively.

### A. Optimal Power Allocation for Co-Mode-Interference-Free Model Based RowComms

For the concentric UCAs without co-mode interference, the capacity, denoted by $C_{\mathrm{CIF}}$, can be derived as follows:

$$\begin{aligned} C_{\mathrm{CIF}} &= \sum_{l=\frac{1-U}{2}}^{U/2} B \log_2 (1 + \mathrm{SNR}_l) \\ &= \sum_{l=\frac{1-U}{2}}^{U/2} B \log_2 \left\{ \det \left[ \boldsymbol{I} + \frac{|\boldsymbol{s}_l|^2}{\sigma_l^2 |(\boldsymbol{H}_l^{\mathrm{H}} \boldsymbol{H}_l)^{-1} \boldsymbol{H}_l^{\mathrm{H}}|^2} \right] \right\} \\ &= \sum_{l=\frac{1-U}{2}}^{U/2} B \log_2 \left\{ \det \left[ \boldsymbol{I} + \frac{\overline{\boldsymbol{H}}_l |\boldsymbol{s}_l|^2 \overline{\boldsymbol{H}}_l^{\mathrm{H}}}{\sigma_l^2} \right] \right\}, \end{aligned} \tag{23}$$

where $\overline{\boldsymbol{H}}_l = \left[(\boldsymbol{H}_l^{\mathrm{H}}\boldsymbol{H}_l)^{-1}\boldsymbol{H}_l^{\mathrm{H}}\right]^{-1}$, $B$ is system bandwidth. We can obtain that $\overline{\boldsymbol{H}}_l = \boldsymbol{Q}\boldsymbol{D}\boldsymbol{W}^{\mathrm{H}}$ through singular value decomposition, where $\boldsymbol{Q}$ and $\boldsymbol{W}^{\mathrm{H}}$ are unitary matrices and $\boldsymbol{D}$ is a diagonal matrix. Then, we can rewrite $\overline{\boldsymbol{H}}_l|\boldsymbol{s}_l|^2\overline{\boldsymbol{H}}_l^{\mathrm{H}}$ as follows:

$$\overline{\boldsymbol{H}}_l|\boldsymbol{s}_l|^2\overline{\boldsymbol{H}}_l^{\mathrm{H}} = \boldsymbol{Q}\boldsymbol{D}\boldsymbol{W}^{\mathrm{H}}|\boldsymbol{s}_l|^2(\boldsymbol{Q}\boldsymbol{D}\boldsymbol{W}^{\mathrm{H}})^{\mathrm{H}} = \begin{bmatrix} P_{1,l}\gamma_{1,l} & 0 & \cdots & 0 \\ 0 & P_{2,l}\gamma_{2,l} & \cdots & 0 \\ \vdots & \vdots & \ddots & \vdots \\ 0 & 0 & \cdots & P_{\mathrm{rank}(\boldsymbol{H}_l),l}\gamma_{\mathrm{rank}(\boldsymbol{H}_l),l} \end{bmatrix}. \tag{24}$$

where $\gamma_{i,l}$ is the square of singular value corresponding to the matrix $\overline{\boldsymbol{H}}_l$. Taking Eq. (24) into Eq. (23), we can obtain the capacity $C_{\mathrm{CIF}}$ as follows:

$$C_{\mathrm{CIF}} = \sum_{l=\frac{1-U}{2}}^{U/2} \sum_{i=1}^{\mathrm{rank}(\boldsymbol{H}_l)} B\log_2\left(1 + \frac{P_{i,l}\gamma_{i,l}}{\sigma_l^2}\right). \tag{25}$$

where $\mathrm{rank}\left(\boldsymbol{H}_l\right)$ denotes the rank of $\boldsymbol{H}_l$ and $P_{i,l}$ denotes the power allocation scheme. We aim to maximize the capacity of co-mode-interference-free model based RowComms. Thus, we formulate the capacity maximization problem, denoted by $\boldsymbol{P1}$, as follows:

$$\boldsymbol{P1}\text{:} \quad \max_{\substack{P_{i,l}: \\ 1\le i\le \mathrm{rank}(\boldsymbol{H}_l), \\ (1-U)/2\le l\le U/2}} \sum_{l=\frac{1-U}{2}}^{U/2} \sum_{i=1}^{\mathrm{rank}(\boldsymbol{H}_l)} B\log_2\left(1 + \frac{P_{i,l}\gamma_{i,l}}{\sigma_l^2}\right) \tag{26}$$

$$\text{s.t.}: \ 1).\ \sum_{l=\frac{1-U}{2}}^{U/2} \sum_{i=1}^{\mathrm{rank}(\boldsymbol{H}_l)} P_{i,l} \le \overline{P}; \tag{27}$$

$$2).\ P_{i,l} \ge 0,\ (1-U)/2 \le l \le U/2,\ 1 \le i \le \mathrm{rank}(\boldsymbol{H}_l), \tag{28}$$

where

$$\boldsymbol{\gamma} = \begin{bmatrix} \gamma_{1,\frac{1-U}{2}} & \cdots & \gamma_{1,0} & \cdots & \gamma_{1,\frac{U}{2}} \\ \gamma_{2,\frac{1-U}{2}} & \cdots & \gamma_{2,0} & \cdots & \gamma_{2,\frac{U}{2}} \\ \vdots & \vdots & \ddots & \cdots & \vdots \\ \gamma_{\mathrm{rank}(\boldsymbol{H}_l),\frac{1-U}{2}} & \cdots & \gamma_{\mathrm{rank}(\boldsymbol{H}_l),0} & \cdots & \gamma_{\mathrm{rank}(\boldsymbol{H}_l),\frac{U}{2}} \end{bmatrix} \tag{29}$$

is the square of singular value matrix for all channels of the concentric UCAs based RowComms and $\overline{P}$ is the average power constraint.

It is clear that $\boldsymbol{P1}$ is a strictly convex optimization problem. To solve $\boldsymbol{P1}$, we construct the Lagrangian function, denoted by $J$, for $\boldsymbol{P1}$ as follows:

$$J = \sum_{l=\frac{1-U}{2}}^{U/2} \sum_{i=1}^{\mathrm{rank}(\boldsymbol{H}_l)} B \log_2\left(1 + \frac{P_{i,l}\gamma_{i,l}}{\sigma_l^2}\right) + \sum_{i=1}^{\mathrm{rank}(\boldsymbol{H}_l)} \sum_{l=\frac{1-U}{2}}^{U/2} \epsilon_{i,l} P_{i,l} - \mu\left(\sum_{l=\frac{1-U}{2}}^{U/2} \sum_{i=1}^{\mathrm{rank}(\boldsymbol{H}_l)} P_{i,l} - \overline{P}\right), \tag{30}$$

where $\mu \geq 0$ and $\epsilon_{i,l} \geq 0$ $\big(1 \leq i \leq \mathrm{rank}(\boldsymbol{H}_l),\ (1-U)/2 \leq l \leq U/2\big)$ are the Lagrangian multipliers associated with the constraints specified by Eqs. (27) and (28), respectively. Taking the derivative for $J$ with the respect to $P_{i,l}$ and setting the derivatives equal to zero, we can obtain a set of $\sum_{l=(1-U)/2}^{U/2} \mathrm{rank}(\boldsymbol{H}_l)$ equations as follows:

$$\frac{\partial J}{\partial P_{i,l}} = \frac{B\gamma_{i,l}}{(\sigma_l^2 + P_{i,l}\gamma_{i,l})\log 2} - \mu + \epsilon_{i,l} = 0, \quad (1-U)/2 \leq l \leq U/2,\ 1 \leq i \leq \mathrm{rank}(\boldsymbol{H}_l). \tag{31}$$

According to the principle of complementary slackness, we have $\epsilon_{i,l}P_{i,l} = 0$ for $\forall l \in [(1-U)/2, U/2]$ and $\forall i \in [1, \mathrm{rank}(\boldsymbol{H}_l)]$. Then, Eq. (31) can be reduced to as follows:

$$\frac{B\gamma_{i,l}}{(\sigma_l^2 + P_{i,l}\gamma_{i,l})\log 2} - \mu = 0. \tag{32}$$

Therefore, we can obtain the power allocation scheme as follows:

$$P_{i,l} = \frac{B}{\mu \log 2} - \frac{\sigma_l^2}{\gamma_{i,l}} \tag{33}$$

where $1 \leq i \leq \mathrm{rank}(\boldsymbol{H}_l)$ and $(1-U)/2 \leq l \leq U/2$.

However, it is possible for obtained $P_{i,l}$ to be less than zero, i.e. not all channels corresponding to all OAM-modes of all transmit UCAs can achieve high capacity transmission. Thus, we define $(P_{i,l})^+$, called the optimal power allocation scheme for $1 \leq i \leq \mathrm{rank}(\boldsymbol{H}_l)$ and $(1-U)/2 \leq l \leq U/2$ as follows:

$$(P_{i,l})^+ = \begin{cases} P_{i,l}, & P_{i,l} > 0; \\ 0, & P_{i,l} < 0. \end{cases} \tag{34}$$

Then, the optimal power allocation scheme can be derived as follows:

$$(P_{i,l})^+ = \begin{cases} \frac{B}{\mu^* \log 2} - \frac{\sigma_l^2}{\gamma_{i,l}}, & \mu^* < \frac{B\gamma_{i,l}}{\sigma_l^2 \log 2}; \\ 0, & \mu^* > \frac{B\gamma_{i,l}}{\sigma_l^2 \log 2}, \end{cases} \tag{35}$$

where $1 \leq i \leq \mathrm{rank}(\boldsymbol{H}_l)$, $(1-U)/2 \leq l \leq U/2$, and $\mu^*$ is the optimal Lagrangian multiplier

**Algorithm 2** : The OAM Power Allocation Scheme

Put $\sigma_l^2/\gamma_{i,l}$ in order and form the corresponding sequence which contains $q$ elements.
$\widetilde{P}=0$
**for** $z=1$ to $(q-1)$ **do**
  $\widetilde{P}=\widetilde{P}+(b_{z+1}-b_z)$
  **if** $\widetilde{P}>\overline{P}$ **then**
    $\mu^*=\frac{Bz}{\log 2\left(zb_{z+1}-\widetilde{P}+\overline{P}\right)}$; break
  **end if**
**end for**
**if** $\widetilde{P}<\overline{P}$ **then**
  $\mu^*=\frac{Bq}{\log 2\left(qb_q-\widetilde{P}+\overline{P}\right)}$
**end if**
The scheme can be obtained as follows:

$$(P_{i,l})^+=\begin{cases}\frac{B}{\mu^*\log 2}-\frac{\sigma_l^2}{\gamma_{i,l}}, & \mu^*<\frac{B\gamma_{i,l}}{\sigma_l^2\log 2};\\ 0, & \mu^*>\frac{B\gamma_{i,l}}{\sigma_l^2\log 2}.\end{cases}$$

corresponding to $\mu$. The optimal value for $\mu^*$ can be numerically obtained by substituting $P_{i,l}$ into

$$\sum_{l=\frac{1-U}{2}}^{U/2}\sum_{i=1}^{\mathrm{rank}(\boldsymbol{H}_l)}P_{i,l}=\overline{P}. \tag{36}$$

Thus, it is of great importance to obtain the optimal Lagrangian multiplier to maximize the capacity. Then, we develop the optimal OAM power allocation scheme, illustrated in **Algorithm 2**, where $b_z$ is the $z$th element in the sequence and $\widetilde{P}$ represents the sum of allocated power.

### *B. Optimal Power Allocation for Co-Mode-Interference-Contained Model Based RowComms*

For concentric UCAs based RowComms with co-mode interference, the capacity, denoted by $C_{\mathrm{CIC}}$, can be derived as follows:

$$\begin{aligned}C_{\mathrm{CIC}} &= \sum_{l=\frac{1-U}{2}}^{U/2}\sum_{i=1}^{N}B\log_2\left(1+\mathrm{SINR}_{i,l}\right)\\ &= \sum_{l=\frac{1-U}{2}}^{U/2}\sum_{i=1}^{N-1}B\log_2\left(1+\frac{\frac{P_{i,l}}{\|(\mathbf{w}_i^l)_{k_i}\|^2}}{\sigma_l^2+\sum\limits_{\varrho\notin\mathcal{K}_i^l}\frac{P_{\varrho,l}}{\|(\mathbf{w}_i^l)_{\varrho}\|^2}}\right)+\sum_{l=\frac{1-U}{2}}^{U/2}\log_2\left(1+\frac{P_{N,l}}{\sigma_l^2\|(\mathbf{w}_N^l)_{k_N}\|^2}\right).\end{aligned} \tag{37}$$

In the following discussion, we denote by $\boldsymbol{\chi}$ the channel power gains for co-mode-interference-contained based RowComms, which is expressed as follows:

$$\boldsymbol{\chi} = \left[\boldsymbol{\chi}^{\frac{1-U}{2}}, \cdots, \boldsymbol{\chi}^{0}, \boldsymbol{\chi}^{1}, \cdots, \boldsymbol{\chi}^{\frac{U}{2}}\right], \tag{38}$$

where $\boldsymbol{\chi}^{l}$ is a matrix given as follows:

$$\boldsymbol{\chi}^{l} = \begin{bmatrix} \frac{1}{\|(\mathbf{w}_1^l)_{k_1}\|^2} & \frac{1}{\|(\mathbf{w}_1^l)_{k_2}\|^2} & \cdots & \frac{1}{\|(\mathbf{w}_1^l)_{k_N}\|^2} \\ \frac{1}{\|(\mathbf{w}_2^l)_{k_1}\|^2} & \frac{1}{\|(\mathbf{w}_2^l)_{k_2}\|^2} & \cdots & \frac{1}{\|(\mathbf{w}_2^l)_{k_N}\|^2} \\ \vdots & \vdots & \ddots & \vdots \\ \frac{1}{\|(\mathbf{w}_N^l)_{k_1}\|^2} & \frac{1}{\|(\mathbf{w}_N^l)_{k_N}\|^2} & \cdots & \frac{1}{\|(\mathbf{w}_N^l)_{k_N}\|^2} \end{bmatrix}. \tag{39}$$

Our goal is to achieve maximum capacity for co-mode-interference-contained based RowComms. Thus, we formulate the capacity maximization problem, denoted by ***P*2**, for co-mode-interference-contained based RowComms as follows:

$$\textbf{\textit{P}2:} \quad \max_{\substack{P_{i,l}: \\ 1 \leq i \leq \mathrm{rank}(\boldsymbol{H}_l), \\ (1-U)/2 \leq l \leq U/2}} \quad C_{\mathrm{CIC}} \tag{40}$$

$$\text{s.t.}: \ 1). \sum_{l=\frac{1-U}{2}}^{U/2} \sum_{i=1}^{N} P_{i,l} \leq \overline{P}; \tag{41}$$

$$2). \ P_{i,l} \geq 0, \ (1-U)/2 \leq l \leq U/2, \ 1 \leq i \leq \mathrm{rank}(\boldsymbol{H}_l), \tag{42}$$

where $\overline{P}$ represents the average power constraint.

Because the values of $\|(\mathbf{w}_i^l)_{k_i}\|^2$ randomly change when the transmission distance changes, it is difficult to directly obtain whether ***P*2** is a convex optimization problem or not. However, since we can approximate $C_{\mathrm{CIC}}$ to $\widetilde{C}_{\mathrm{CIC}}$, which is given as follows:

$$\widetilde{C}_{\mathrm{CIC}} = \sum_{l=\frac{1-U}{2}}^{U/2} \sum_{i=1}^{N} B \log_2 \left( \frac{\frac{P_{i,l}}{\|(\mathbf{w}_i^l)_{k_i}\|^2}}{\sigma_l^2 + \frac{P_{i+1,l}}{\|(\mathbf{w}_i^l)_{k_{i+1}}\|^2}} \right) + \sum_{l=\frac{1-U}{2}}^{U/2} B \log_2 \left( \frac{P_{N,l}}{\sigma_l^2 \|(\mathbf{w}_N^l)_{k_N}\|^2} \right), \tag{43}$$

problem ***P*2** can be converted to problem ***P*3** as follows:

$$\textbf{\textit{P}3:} \quad \max_{\substack{P_{i,l}:\\ 1\leq i\leq \mathrm{rank}(\boldsymbol{H}_l),\\ (1-U)/2\leq l\leq U/2}} \sum_{l=\frac{1-U}{2}}^{U/2}\sum_{i=1}^{N} B\log_2\left(\frac{\frac{P_{i,l}}{\|(\mathbf{w}_i^l)_{k_i}\|^2}}{\sigma_l^2+\frac{P_{i+1,l}}{\|(\mathbf{w}_i^l)_{k_{i+1}}\|^2}}\right)+\sum_{l=\frac{1-U}{2}}^{U/2} B\log_2\left(\frac{P_{N,l}}{\sigma_l^2\|(\mathbf{w}_N^l)_{k_N}\|^2}\right). \tag{44}$$

$$\text{s.t.}: \ 1).\ \sum_{l=\frac{1-U}{2}}^{U/2}\sum_{i=1}^{N} P_{i,l}\leq \overline{P}; \tag{45}$$

$$2).\ P_{i,l}\geq 0,\ (1-U)/2\leq l\leq U/2,\ 1\leq i\leq \mathrm{rank}(\boldsymbol{H}_l). \tag{46}$$

The following Lemma 1 shows the convexity of problem ***P*3**.

*Lemma 1:* The problem ***P*3** is a strictly convex optimization problem.

*Proof:* For any OAM-mode, when $i=1$, we define the function $f_1(P_{i,l})$, which is given as follows:

$$f_1(P_{1,l})=\log_2\left(\frac{\frac{P_{1,l}}{\|(\mathbf{w}_1^l)_{k_1}\|^2}}{\sigma_l^2+\frac{P_{2,l}}{\|(\mathbf{w}_1^l)_{k_2}\|^2}}\right). \tag{47}$$

Because $\log_2(\cdot)$ is a monotonically increasing function, $f_1(P_{1,l})$ monotonically increases as $P_{1,l}$ increases.

For any OAM-mode, when $1<i\leq N$, we define the function $f_2(P_{i,l})$ as follows:

$$f_2(P_{i,l})=\log_2\left(\frac{\frac{P_{i-1,l}}{\|(\mathbf{w}_{i-1}^l)_{k_{i-1}}\|^2}}{\sigma_l^2+\frac{P_{i,l}}{\|(\mathbf{w}_{i-1}^l)_{k_i}\|^2}}\right)+\log_2\left(\frac{\frac{P_{i,l}}{\|(\mathbf{w}_i^l)_{k_i}\|^2}}{\sigma_l^2+\frac{P_{i+1,l}}{\|(\mathbf{w}_i^l)_{k_{i+1}}\|^2}}\right). \tag{48}$$

For the sake of convenience in writing, we can simplify Eq. (48) as follows:

$$f_2(P_{i,l})=\log_2\left(\frac{A}{B+P_{i,l}}\right)+\log_2\left(\frac{P_{i,l}}{C}\right)=\log_2\left(\frac{AP_{i,l}}{BC+CP_{i,l}}\right), \tag{49}$$

where $A, B,$ and $C$, given by

$$\begin{cases} A=\frac{P_{i-1,l}\|(\mathbf{w}_{i-1}^l)_{k_i}\|^2}{\|(\mathbf{w}_{i-1}^l)_{k_{i-1}}\|^2}; \\ B=\sigma_l^2\|(\mathbf{w}_{i-1}^l)_{k_i}\|^2; \\ C=\sigma_l^2\|(\mathbf{w}_i^l)_{k_i}\|^2+\frac{P_{i+1,l}\|(\mathbf{w}_i^l)_{k_i}\|^2}{\|(\mathbf{w}_i^l)_{k_{i+1}}\|^2}, \end{cases} \tag{50}$$

are larger than zero. It is clear that the function $AP_{i,l}/(BC+CP_{i,l})$ monotonically increases as $P_{i,l}$ increases. Thus, the function $f_2(P_{i,l})$ is a monotonically increasing function with respect to $P_{i,l}$.

In summary, for any $P_{i,l}$, the function $\widetilde{C}_{\rm CIC}$ monotonically increases as $P_{i,l}$ increases. Then, the function $\widetilde{C}_{\rm CIC}$ is strictly concave on the space. It is clear to verify that Eqs. (45) and (46) are linear on the space spanned by $P_{i,l}$. Thus, problem ***P*3** is a strictly convex optimization problem. ■

Since problem ***P*3** is a strictly convex optimization, we can construct the Lagrangian function, denoted by $J_2$, for problem ***P*3** as follows:

$$J_2 = \sum_{l=\frac{1-U}{2}}^{U/2} \sum_{i=1}^{N} B \log_2 \left( \frac{\frac{P_{i,l}}{\|(\mathbf{w}_i^l)_{k_i}\|^2}}{\sigma_l^2 + \frac{P_{i+1,l}}{\|(\mathbf{w}_i^l)_{k_{i+1}}\|^2}} \right) + \sum_{l=\frac{1-U}{2}}^{U/2} B \log_2 \left( \frac{P_{N,l}}{\sigma_l^2 \|(\mathbf{w}_N^l)_{k_N}\|^2} \right) + \sum_{l=\frac{1-U}{2}}^{U/2} \sum_{i=1}^{N} \varsigma_{i,l} P_{i,l} - \nu \left( \sum_{l=\frac{1-U}{2}}^{U/2} \sum_{i=1}^{N} P_{i,l} - \overline{P} \right), \tag{51}$$

where $\varsigma_{i,l}$ and $\nu$ are the Lagrangian multipliers associated with the constraints specified by Eqs. (45) and (46), respectively. According to the principle of complementary slackness, we can obtain that $\varsigma_{i,l} P_{i,l}$ is equal to zero. Then, we can take the derivatives for $J_2$ with respect to $P_{i,l}$ as follows:

$$\begin{cases} \frac{\partial J_2}{\partial P_{1,l}} = \frac{B}{P_{1,l} \log 2} - \nu = 0, & n = 1; \\ \frac{\partial J_2}{\partial P_{i,l}} = \frac{B}{P_{i,l} \log 2} - \frac{B}{\sigma_l^2 \|(\mathbf{w}_{i-1}^l)_{k_i}\|^2 \log 2 + P_{i,l} \log 2} - \nu = 0, & 1 < n \leq N. \end{cases} \tag{52}$$

Then, associated with Eq. (52), we can derive the optimal power allocations as follows:

$$\begin{cases} P_{1,l} = \frac{B}{\nu^* \log 2}, & n = 1; \\ P_{i,l} = \frac{1}{2} \left( \sqrt{\sigma_l^4 \|(\mathbf{w}_{i-1}^l)_{k_i}\|^4 + \frac{4B\sigma_l^2 \|(\mathbf{w}_{i-1}^l)_{k_i}\|^2}{\nu^* \log 2}} - \sigma_l^2 \|(\mathbf{w}_{i-1}^l)_{k_i}\|^2 \right), & 1 < n \leq N, \end{cases} \tag{53}$$

where $\nu^*$ is the optimal Lagrangian multiplier of problem ***P*3** and $\nu^*$ can be numerically derived by taking $P_{i,l}$, specified in Eq. (53), into

$$\sum_{l=\frac{1-U}{2}}^{U/2} \sum_{i=1}^{N} P_{i,l} = \overline{P}. \tag{54}$$

## V. Performance Evaluations

In this section, we evaluate the performance of our developed mode-decomposition schemes and the optimal power allocation schemes for co-mode-interference-free and co-mode-interference-contained models based RowComms. First, we use HFSS to simulate some directional diagrams corresponding to different OAM-modes in singular UCA based RowComms system, where we employ the carrier frequency 2.5 GHz. Then, we evaluate the capacity for different OAM-modes. Finally, we compare the capacity of concentric UCAs based RowComms with the capacity of the singular UCA based RowComms, where we set the system bandwidth as 20 MHz.

Using HFSS, we evaluate the design for the UCA antenna and the isolation coefficient of two adjacent concentric UCAs. Figs. 4, 5 and 6 evaluate the effectiveness of our work.

Figure 3 shows an example for singular UCA to concentric UCAs transformation. In our proposed concentric UCAs based RowComms model, we assume zero mutual coupling among array-elements within one UCA antenna. This is due to the reduced number of array-elements within one UCA antenna of concentric UCAs based RowComms. The distance between two adjacent array-elements within one UCA antenna may increase. Also, mutual coupling among different UCAs may not be increased but even reduced because the smallest distance between two array-elements of adjacent UCAs may be increased. As shown in Fig. 3, we transform the singular UCA equipped with 16 array-elements into two concentric UCAs with 8 array-elements corresponding to each UCA. We can find that the distances between adjacent array-elements corresponding to the two concentric UCAs ( $d_1$ and $d_2$) are both larger than that corresponding to the singular UCA ($d$). Also, the smallest distance between two array-elements of adjacent UCAs ($d_3$) in the concentric UCAs scenario is larger than $d$. Thus, there exists the designed case that we can assume no mutual coupling in the concentric UCAs based RowComms model.

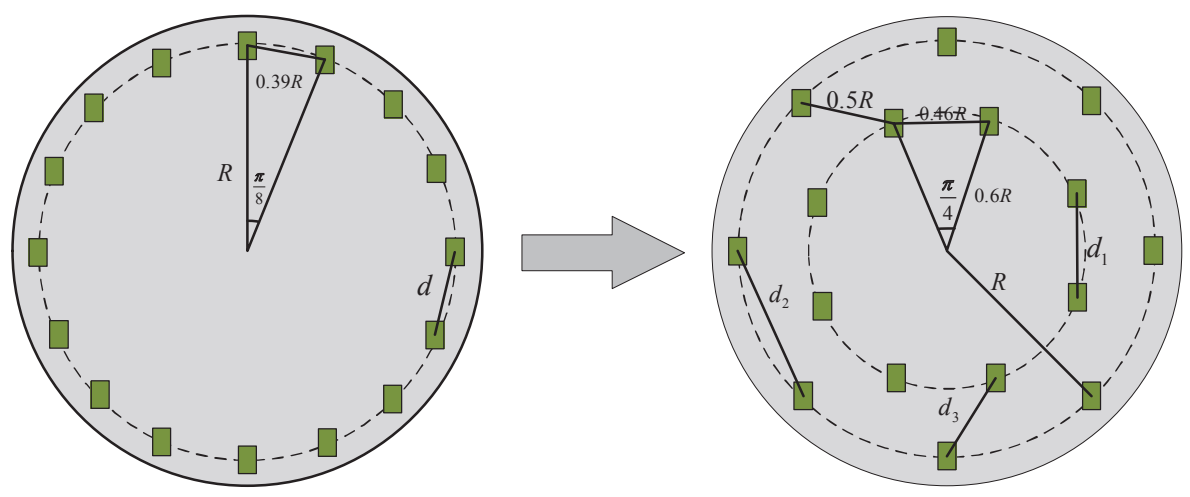

Fig. 3. The example for singular UCA to concentric UCAs transformation.

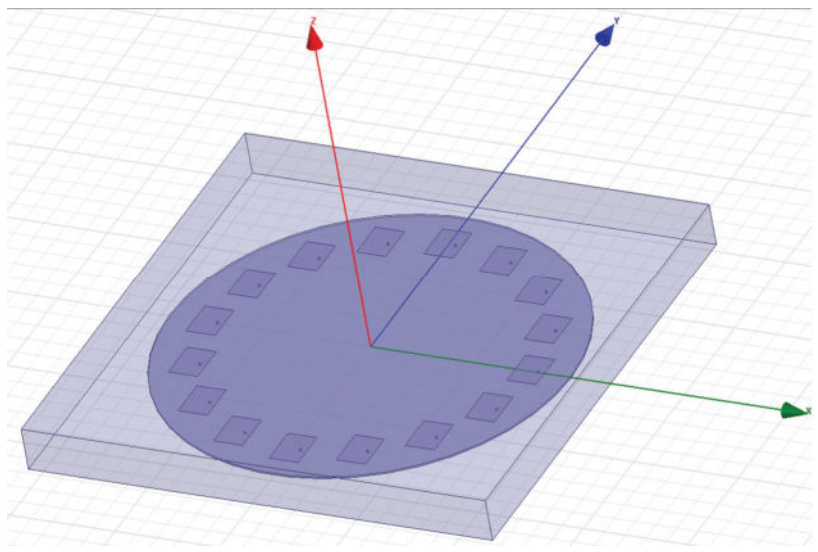

Fig. 4. The singular UCA antenna equipped with 16 array-elements.

Figure 4 displays the UCA antenna which can generate multiple OAM waves with different OAM-modes. The radius of UCA is 192 mm. Referring to [38], we set the thickness as 1.6 mm. As shown in Fig. 4, there are 16 array-elements and the signals on all array-elements have the same amplitude but different azimuthal angles.

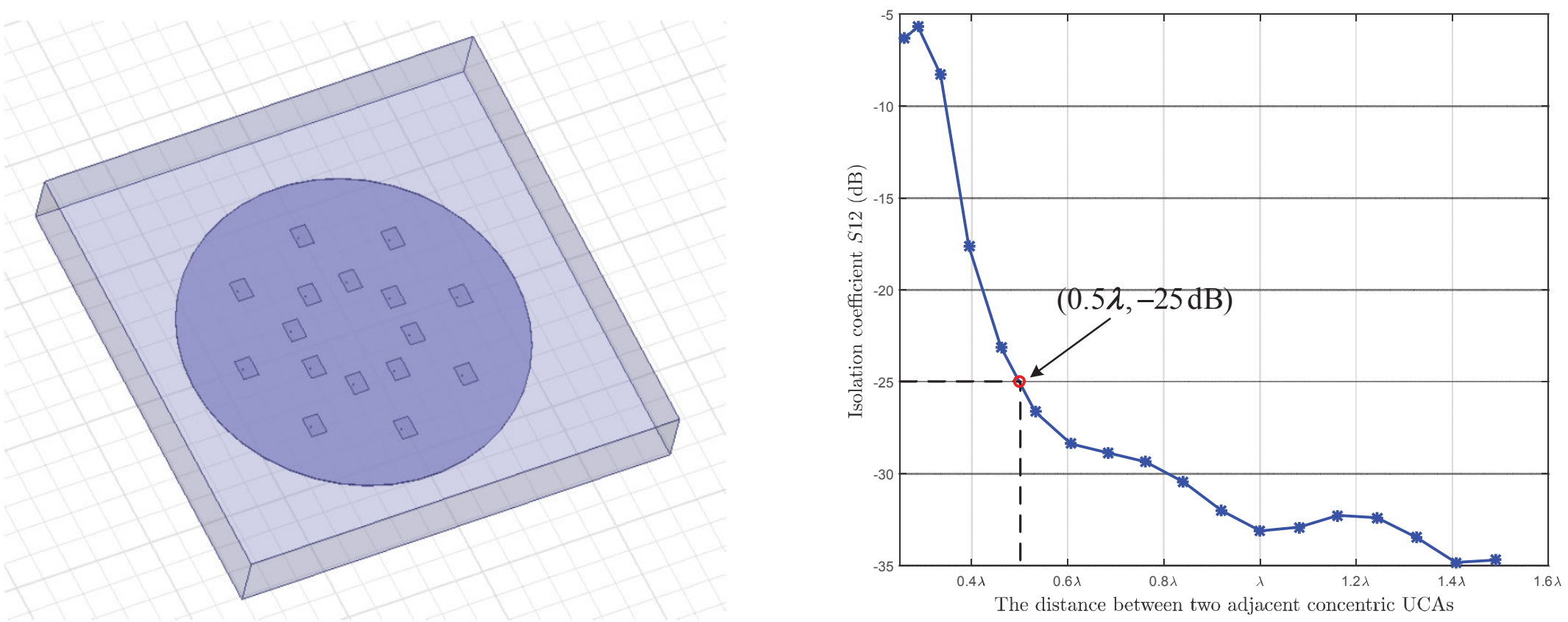

Fig. 5. Concentric UCA antennas and the isolation coefficient $S12$ of two adjacent concentric UCAs.

Figure 5 shows the concentric UCA antennas and the isolation coefficient $S12$ of two adjacent concentric UCAs. We simulate the isolation coefficient $S12$ between two array-elements of adjacent concentric UCAs, reflecting the isolation between the adjacent concentric UCAs. When $S12$ is larger than -25 dB, i.e., the distance between the adjacent concentric UCAs is smaller than 0.5 $\lambda$, the concentric UCAs are correlated and our proposed co-mode-interference-contained model can be applied to this case. When $S12$ is smaller than -25 dB, i.e., the distance between the adjacent concentric UCAs is larger than 0.5 $\lambda$, it is considered that the correlation is very small between the adjacent concentric UCAs and our proposed co-mode-interference-free model can be applied to this scenario.

Figure 6 shows the directional diagrams and phase profiles of OAM waves for different OAM-modes ($l = 0, 1, 2, 4$), which are generated by the UCA antenna given in Fig. 4. As illustrated in Fig. 6, the central null increases as the order of OAM-mode increases. The gain decreases as the order of OAM-mode increases. As a result, the gain of high-order OAM-mode is much less than that of the low-order OAM-mode. That is the reason why the SNRs of high-order OAM-modes are relatively low and thus it is very difficult to utilize the high-order OAM-modes for high capacity in wireless communications.

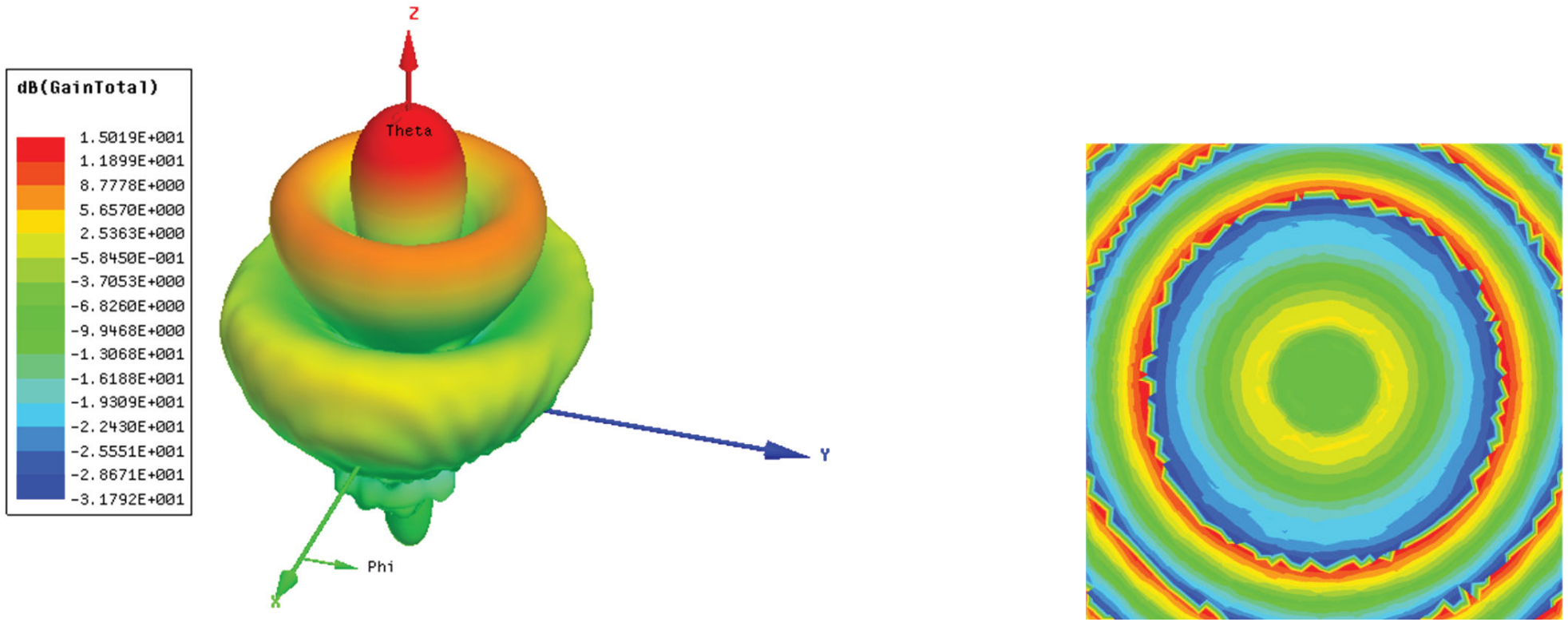

(a) Radiation diagram and phase profile of OAM-mode 0.

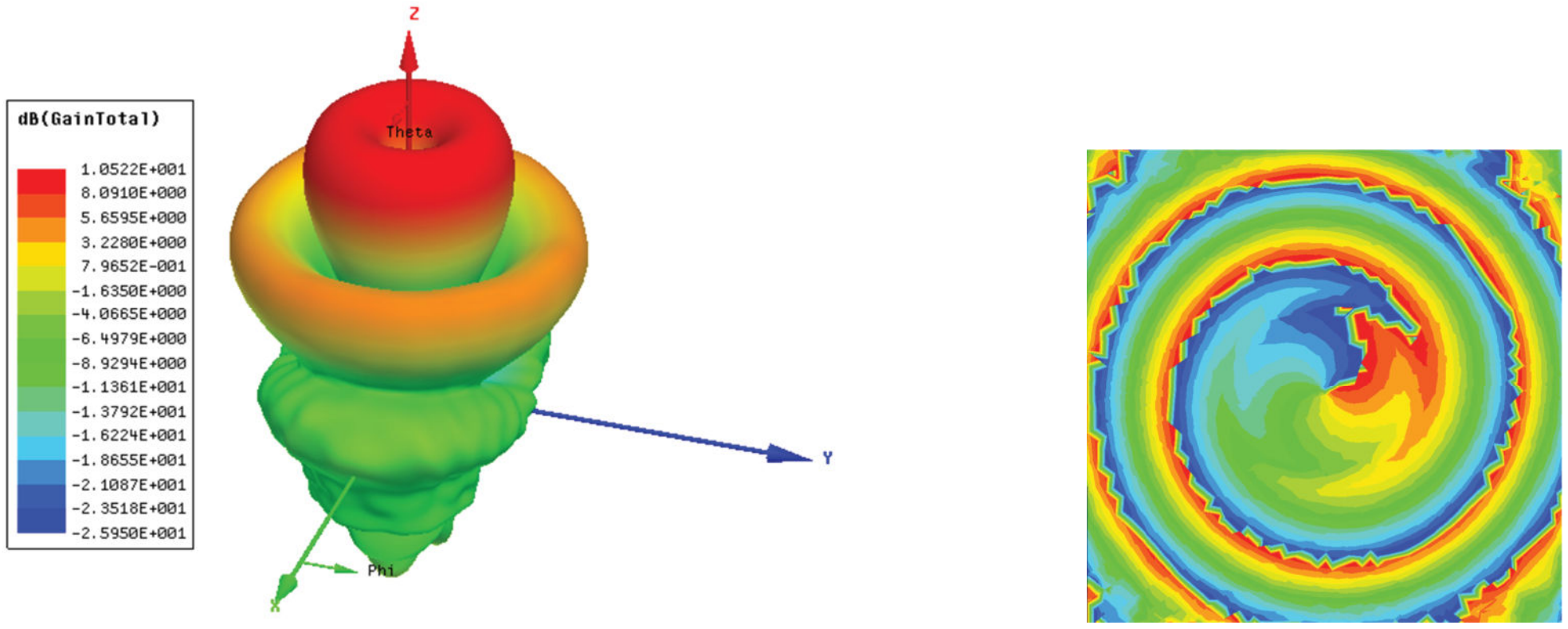

(b) Radiation diagram and phase profile of OAM-mode 1.

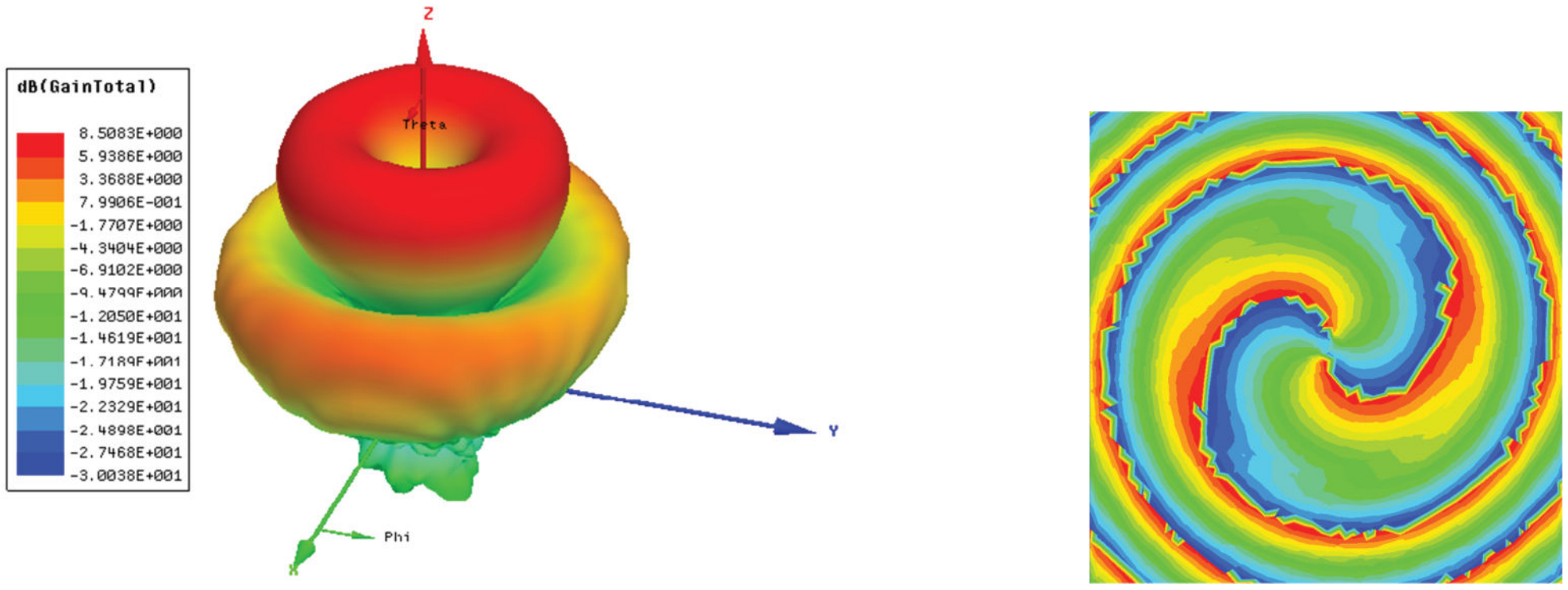

(c) Radiation diagram and phase profile of OAM-mode 2.

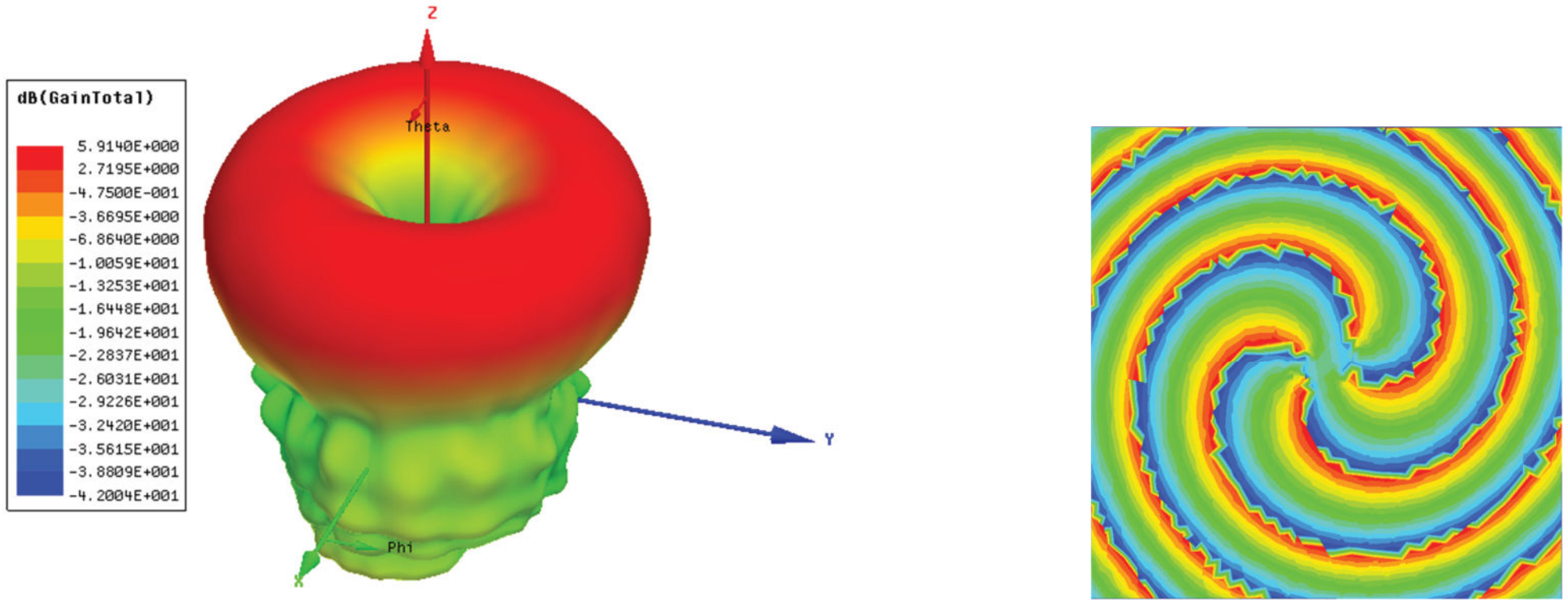

(d) Radiation diagram and phase profile of OAM-mode 4.

Fig. 6. Radiation diagrams and phase profiles for OAM waves with OAM-modes 0, 1, 2, and 4 ($N = M = 1$ and $U = V = 16$).

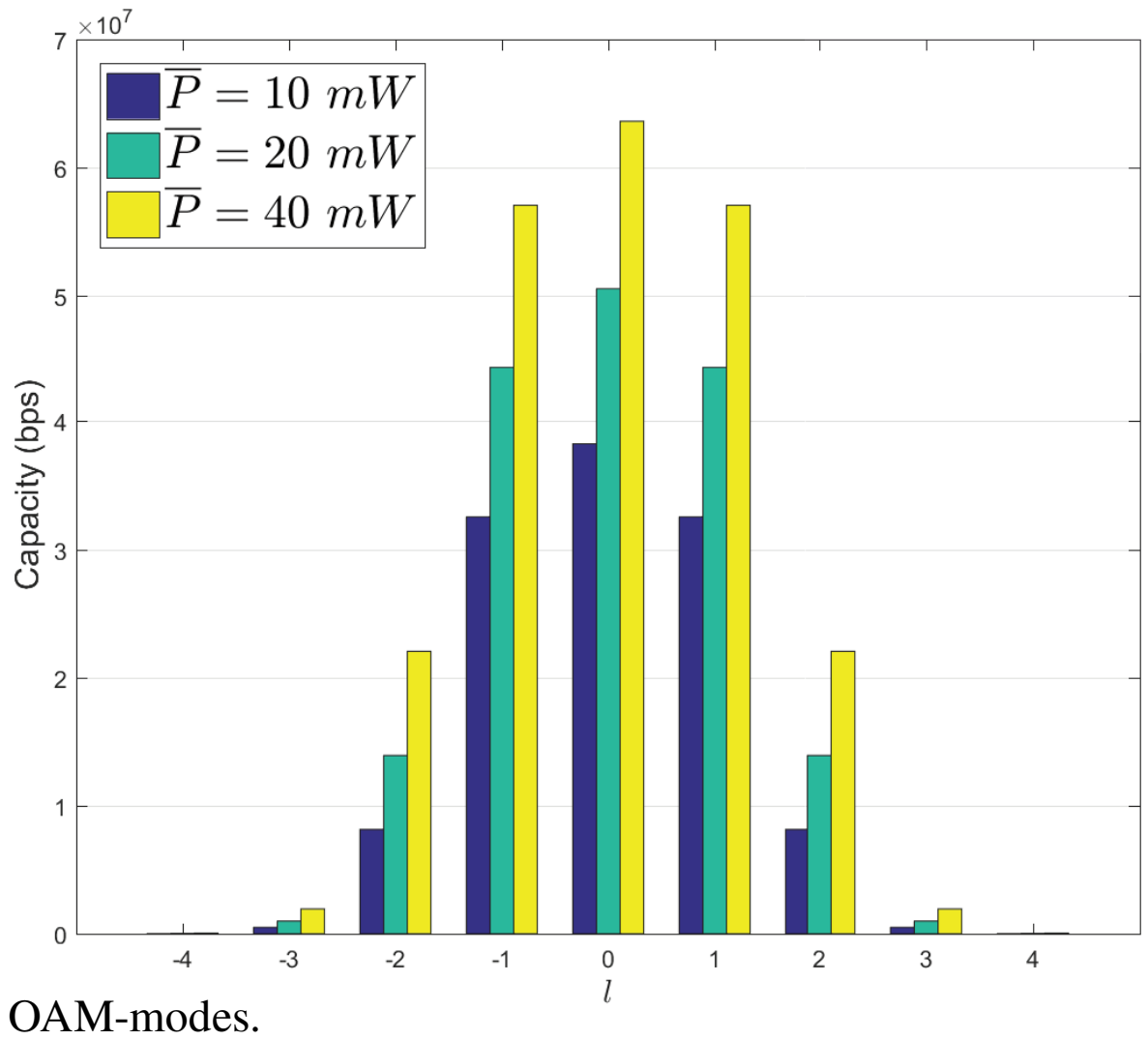

Fig. 7. Capacities of different OAM-modes.

Figure 7 plots the capacities of different OAM-modes for singular UCA based RowComms, where $\overline{P}$ is set as 10 mW, 20 mW, and 40 mW, respectively. As illustrated in Fig. 7, the capacity decreases as the absolute value of OAM-mode number increases. For example, the capacity of OAM-mode 4 is very close to zero. This also validates that the high-order OAM-modes cannot be directly used to achieve high capacity for RowComms. The capacity corresponding to OAM-mode 0 is the largest among the OAM-modes because the OAM wave with OAM-mode 0 is the traditional PE wave, which has relatively small attenuation. Also, we can observe that there is no central null for the OAM wave with OAM-mode 0.

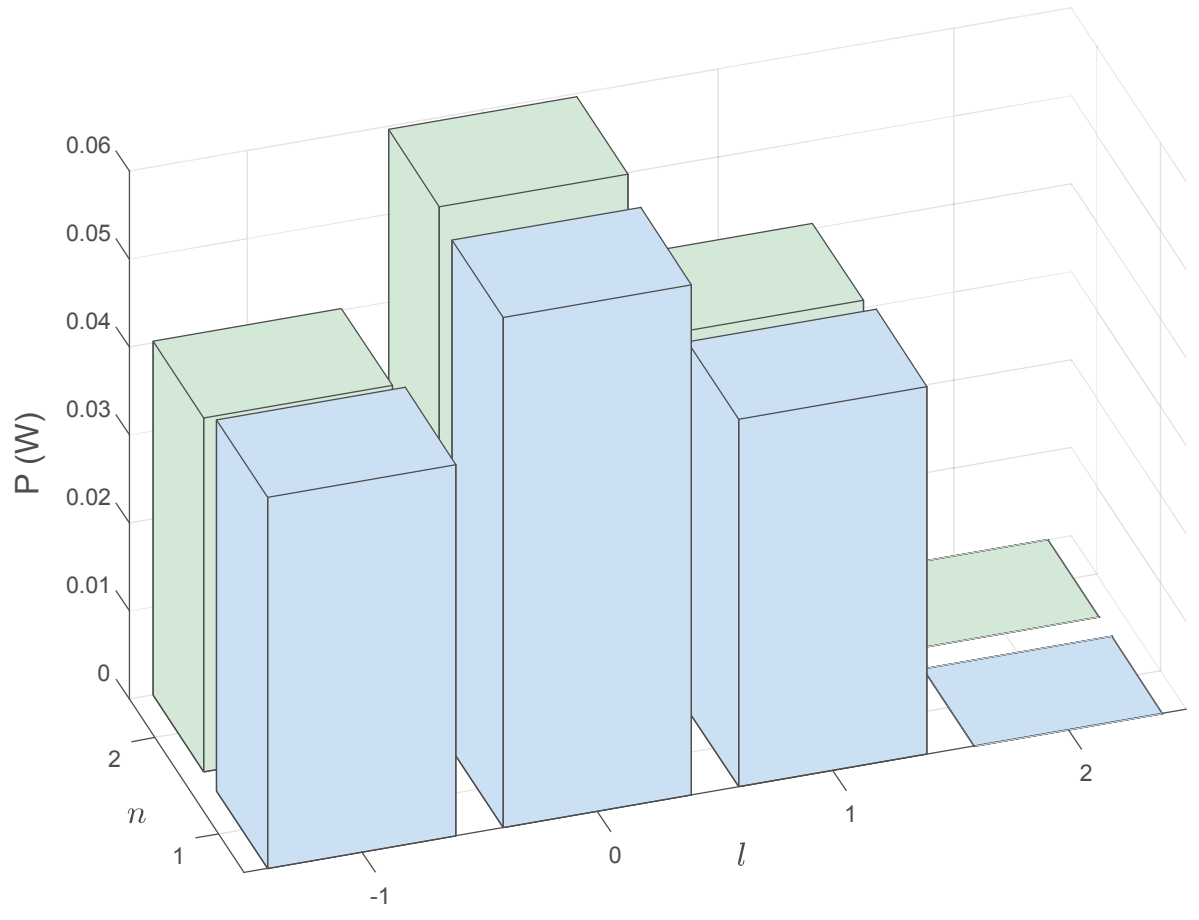

Fig. 8. Power allocation for the concentric UCAs based RowComms without co-mode interference ($N = M = 2$, $U = V = 4$).

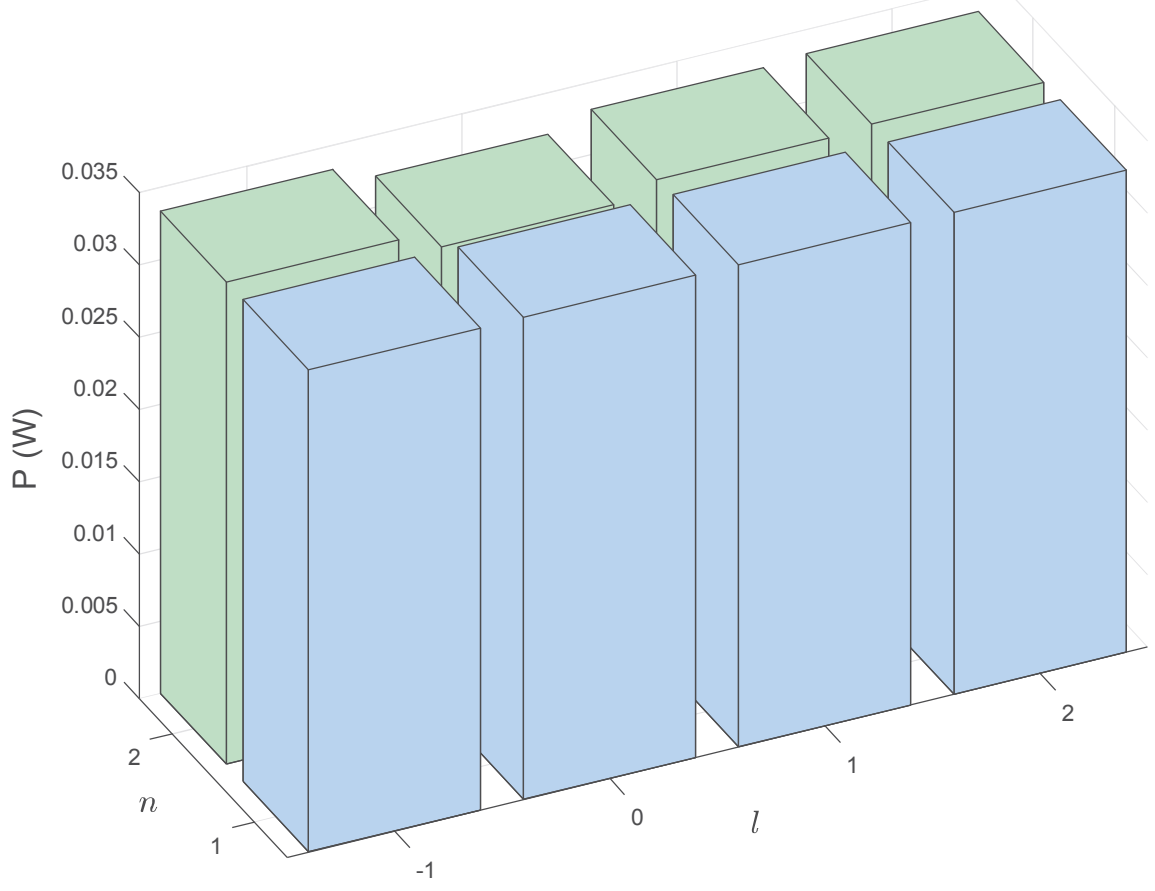

Fig. 9. Power allocation for the concentric UCAs based RowComms with co-mode interference ($N = M = 2$, $U = V = 4$).

Figures 8 and 9 show the power allocation schemes for co-mode-interference-free and co-mode-interference-contained models based RowComms, respectively, where there are 2 concentric UCAs with 4 array-elements on each UCA and the total power is 0.2 W. In Fig. 8, we can observe that the power corresponding to each OAM-mode decreases as the absolute value of OAM-mode number increases for each concentric UCA. The reason why the power of OAM-mode 2 is zero because the attenuation of OAM-mode 2 is relatively larger than those of OAM-modes 0 and 1. The power of the same OAM-modes corresponding to 2 concentric UCAs are nearly the same. This is because there is no co-mode interference between these two concentric UCAs. In Fig. 9, we can observe that the power corresponding to all OAM-modes of each UCA are almost the same. This is because there exists a upper-bound for the power allocated to different OAM-modes of the concentric UCAs based RowComms with co-mode interference.

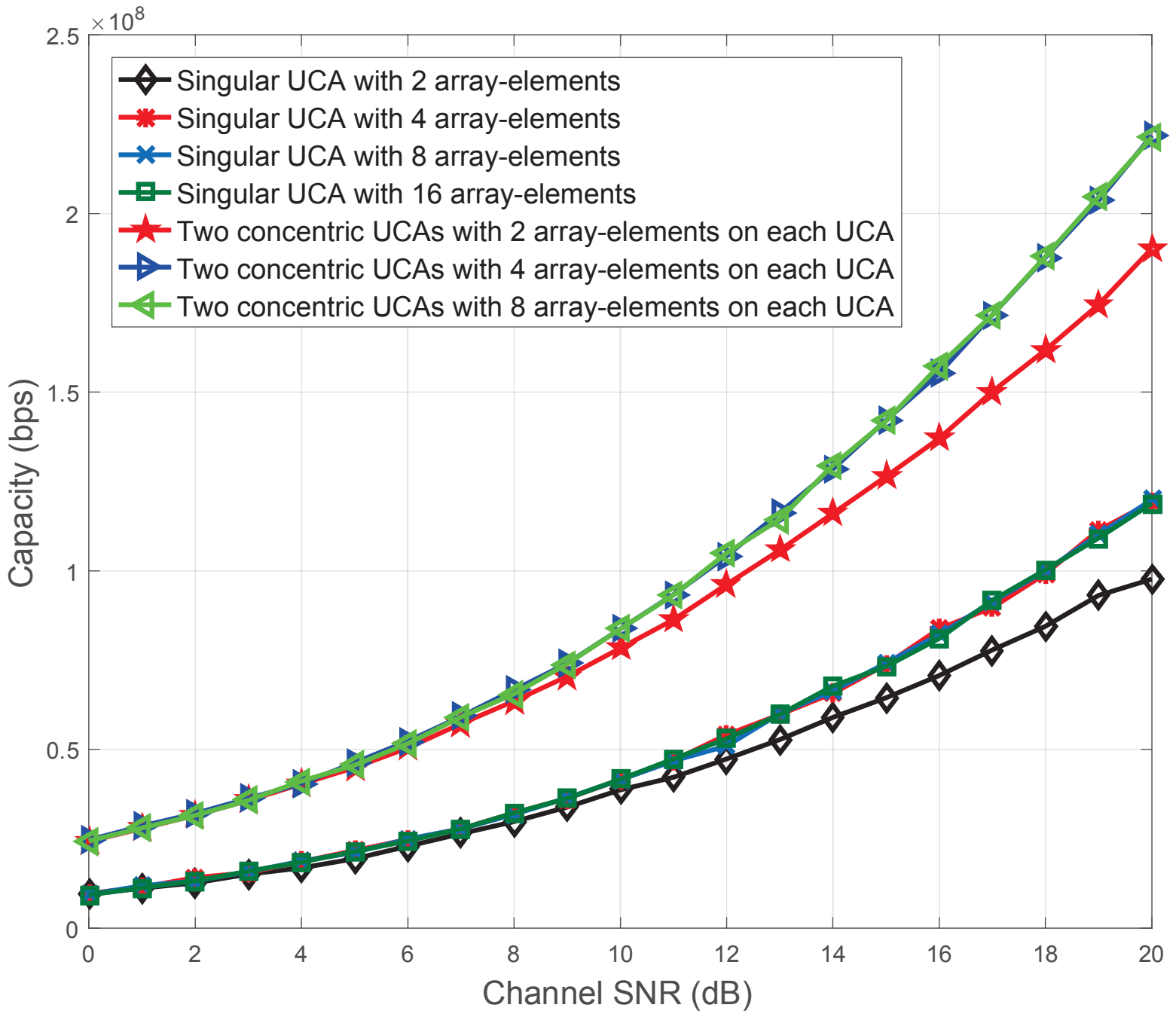

Fig. 10. Capacities of singular UCA based RowComms and the corresponding concentric UCAs based RowComms without co-mode interference.

Figures 10 and 11 depict the capacities of singular UCA and the corresponding concentric UCAs based RowComms without and with co-mode interference, respectively. The number of UCAs at the transmitter and the number of array-elements in each transmit UCA are set the same as the number of UCAs at the receiver and the number of array-elements in each receive UCA, respectively. There are three cases in Figs. 10 and 11: 1). Singular UCA with 16 array-elements and its corresponding 2 concentric UCAs with 8 array-elements on each UCA; 2). Singular UCA with 8 array-elements and its corresponding 2 concentric UCAs with 4 array-elements on each UCA; 3). Singular UCA with 4 array-elements and its corresponding 2 concentric UCAs with 2 array-elements on each UCA. We can observe that the capacity of singular UCA based RowComms is less than that of 2 concentric UCAs based RowComms when the total number of array-elements is the same as shown in Figs. 10 and 11, for co-mode-interference-free and co-mode-interference-contained models based RowComms. This is because the number of low-order OAM-modes is doubled as compared with the number of the corresponding high-order OAM-modes. Based on the four plots corresponding to the capacities of singular UCA based RowComms in Figs. 10 and 11, we can obtain that the capacities keep nearly unchanged as the OAM-mode number increases. That is because the high-order OAM-modes experience severe attenuations, resulting in low SNR for high-order OAM-modes. This is also the reason why two plots corresponding to the capacities of 2 concentric UCAs with 4 array-elements on each UCA and 8 array-elements on each UCA are almost the same. On the other hand, we can observe that the capacities of all singular UCA based RowComms are the same and the capacities of all concentric UCAs based RowComms are also the same when the channel SINR is relatively low. This is because if the OAM beams are not converged, only the low-order OAM-modes can significantly increase the capacity for RowComms. What's more, the gap between the capacities for the singular UCA and the corresponding 2 concentric UCAs based RowComms without co-mode interference is larger than the gap between the capacities for the singular UCA and the corresponding 2 concentric UCAs based RowComms with co-mode interference. This is because the co-mode interference leads to the reduction in the received SNR. Thus, when the size of singular UCA antenna is relatively large, it is suggested to convert the singular UCA to the concentric UCAs based RowComms without co-mode interference.

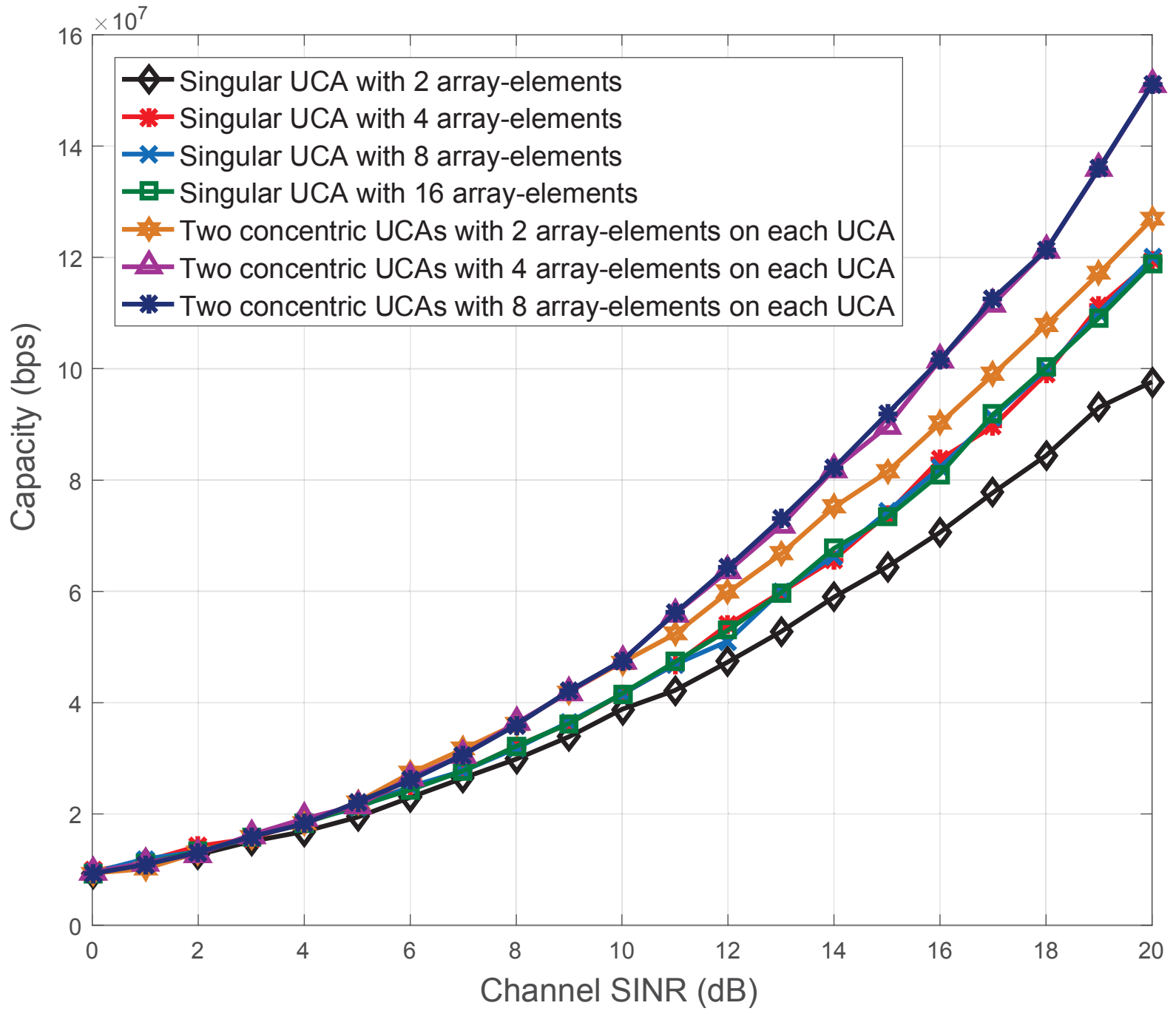

Fig. 11. Capacities of singular UCA based RowComms and the corresponding concentric UCAs based RowComms with co-mode interference.

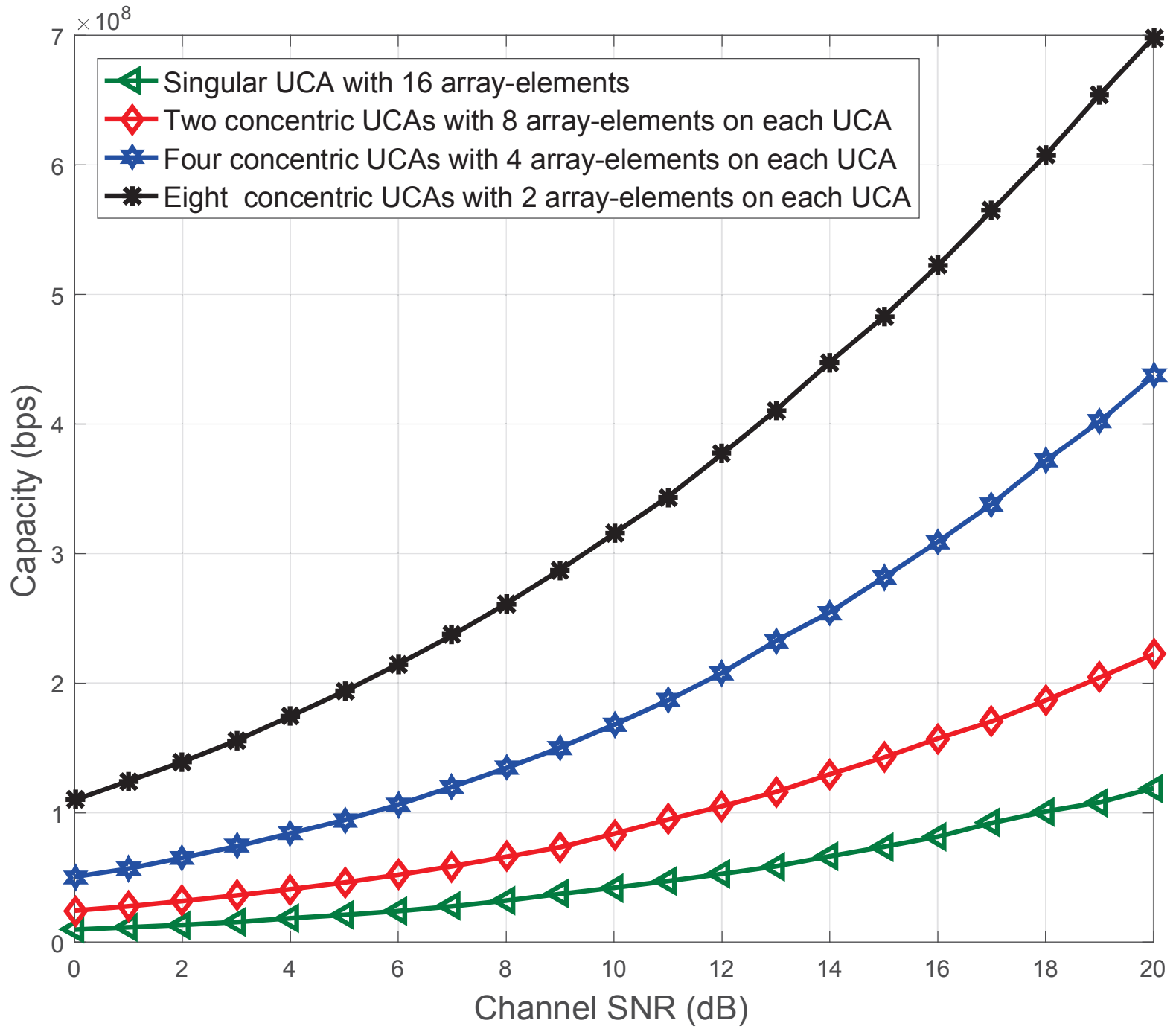

Fig. 12. Capacities of co-mode-interference-free model based RowComms with different number of concentric UCAs.

Figures 12 and 13 plot the capacities of co-mode-interference-free and co-mode-interference-contained models based RowComms versus different number of concentric UCAs. The number of concentric UCAs depends on the diameter and it is not required that the diameter must be rather large. Based on Figs. 12 and 13, we can observe that the capacities of concentric UCAs based RowComms for two cases increase as the number of concentric UCAs increases when the total number of array-elements keeps unchanged. This is because the capacity of concentric UCAs based RowComms increases as the number of low-order modes increases. However, as shown in Fig. 13, we can observe that when the channel SINR is very large, the capacity of four concentric UCAs with 4 array-elements on each UCA is larger than that of 8 concentric UCAs with 2 array-elements on each UCA. This is because the number of co-interfered low-order OAM-modes for the 8 concentric UCAs with 2 array-elements on each UCA is larger than that for 4 concentric UCAs with 4 array-elements on each UCA. Thus, it is highly suggested to fully use the area around the center of transmit UCA and receive UCA for high capacity in RowComms.

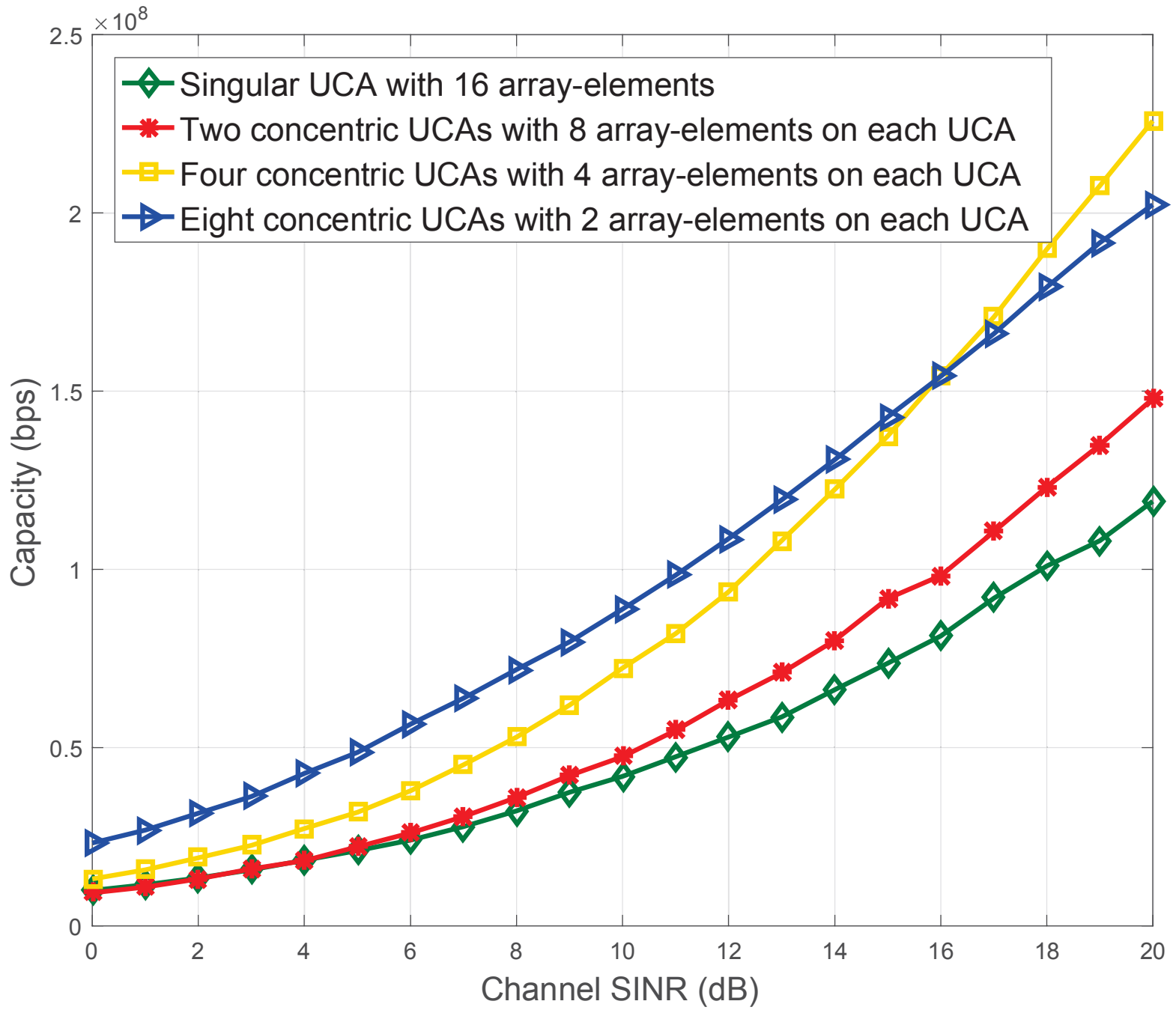

Fig. 13. Capacities of co-mode-interference-contained model based RowComms with different number of concentric UCAs.

## VI. Conclusions

We proposed a concentric UCAs model based RowComms to significantly increase the capacity of RowComms. In particular, we transformed the singular UCA based RowComm into an equivalent concentric UCAs based low-order RowComm for transmitter-receiver aligned scenario. We proposed the co-mode-interference-contained and the co-mode-interference-free models. Then, we developed a mode-decomposition scheme and multiplexing-detection scheme for co-mode-interference-free model and a mode-decomposition scheme and CM-ZF-SIC algo-rithm for co-mode-interference-contained model to recover the transmit signals. The optimal power allocation schemes were also developed to achieve the maximum capacity for concentric UCAs based RowComms. Simulation results validated the mode-decomposition scheme, the multiplexing-detection scheme, the CM-ZF-SIC algorithm, and the optimal power allocation schemes, showing that the concentric UCAs based RowComms can significantly increase the capacity over the singular UCA based RowComms. Future work regarding the transmitter and receiver off-aligned scenario is challenging but important. In practical applications, it is more common for the off-aligned scenario as compared with the aligned scenario. Also, it is difficult to keep the transmit UCA and receive UCA well aligned.

# BIOGRAPHIES

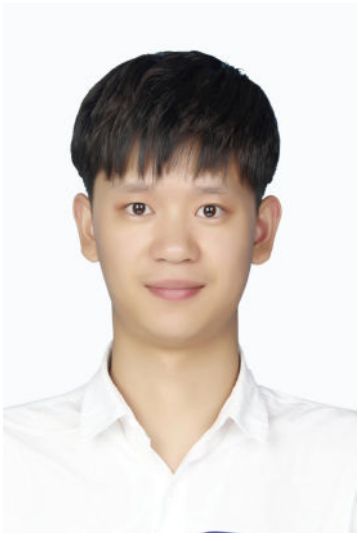

**Haiyue Jing** received the B.S. degree in Telecommunication Engineering from Xidian University, China, in 2017. He is currently pursuing the Ph.D. degree in Telecommunication Engineering with Xidian University. His research interests focus on the generation of orbital-angular-momentum waves, mode-detection schemes, and high capacity of radio vortex wireless communications. (hyjing@stu.xidian.edu.cn)

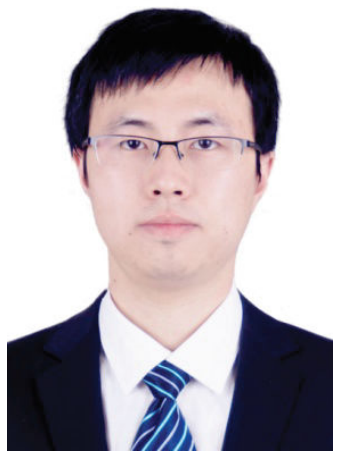

**Wenchi Cheng** (M'14-SM'18) received the B.S. and Ph.D. degrees in telecommunication engineering from Xidian University, China, in 2008 and 2014, respectively, where he is an associate professor. He joined the Department of Telecommunication Engineering, Xidian University, in 2013, as an assistant professor. He worked as a visiting scholar at the Networking and Information Systems Laboratory, Department of Electrical and Computer Engineering, Texas A&M University, College Station, Texas, USA, from 2010 to 2011. His current research interests include 5G wireless networks and orbital-angular-momentum based wireless communications. He has published more than 70 international journal and conference papers in IEEE Journal on Selected Areas in Communications, IEEE Magazines, IEEE INFOCOM, GLOBECOM, and ICC, etc. He received the Young Elite Scientist Award of CAST, the Best Dissertation (Rank 1) of China Institute of Communications, the Best Paper Award for IEEE/CIC ICCC 2018, the Best Paper Nomination for IEEE GLOBECOM 2014, and the Outstanding Contribution Award for Xidian University. He has served or serving as the Associate Editor for IEEE Access, the IoT Session Chair for IEEE 5G Roadmap, the Publicity Chair for IEEE ICC 2019, the Next Generation Networks Symposium Chair for IEEE ICCC 2019, the Workshop Chair for IEEE ICC 2019 Workshop on Intelligent Wireless Emergency Communications Networks, the Workshop Chair for IEEE ICCC 2017 Workshop on Internet of Things. (wccheng@xidian.edu.cn)

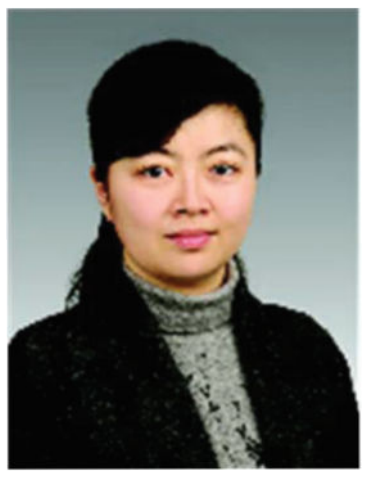

**Zan Li** (SM'14) received B.S. degree in communications engineering in 1998 and the M.S. and Ph.D. degrees in communication and information systems from Xidian University, Xian, China, in 2001 and 2004, respectively. She is currently a Professor with the State Key Laboratory of Integrated Services Networks, School of Telecommunications Engineering, Xidian University. Her research interests include topics on wireless communications and signal processing, such as weak signal detection, spectrum sensing, and cooperative communications. (zanli@xidian.edu.cn)

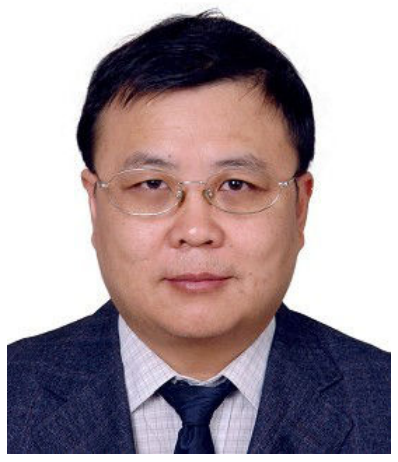

**Hailin Zhang** (M'97) received B.S. and M.S. degrees from Northwestern Polytechnic University, Xian, China, in 1985 and 1988 respectively, and the Ph.D. from Xidian University, Xian, China, in 1991. In 1991, he joined School of Telecommunications Engineering, Xidian University, where he is a senior Professor and the Dean of this school. He is also currently the Director of Key Laboratory in Wireless Communications Sponsored by China Ministry of Information Technology, a key member of State Key Laboratory of Integrated Services Networks, one of the state government specially compensated scientists and engineers, a field leader in Telecommunications and Information Systems in Xidian University, an Associate Director of National 111 Project. Dr. Zhangs current research interests include key transmission technologies and standards on broadband wireless communications for 5G and 5G-beyond wireless access systems. He has published more than 150 papers in journals and conferences. (hlzhang@xidian.edu.cn)